\documentclass[twocolumn]{aastex631}

\newcommand\Msun{\; {\rm M}_{\odot}}
\newcommand\kms{\; {\rm km}\;{\rm s}^{-1}}

\newcommand\kpc{\;{\rm kpc}}

\newcommand\Gyr{\;{\rm Gyr}}

\newcommand\QT{Q_{T}}
\newcommand\QTmin{{Q_{T,\text{min}}}}

\usepackage{amsmath}
\usepackage{graphicx}

\usepackage[nameinlink,capitalise,noabbrev]{cleveref}

\shorttitle{Effects of CMC on Bar Formation}
\shortauthors{JANG \& KIM}

\begin{document}

\title{Effects of the Central Mass Concentration on Bar Formation in Disk Galaxies}

\correspondingauthor{Woong-Tae Kim}
\email{unitree@snu.ac.kr}

\author[0000-0002-7202-4373]{Dajeong Jang}
\affiliation{Department of Physics $\&$ Astronomy, Seoul National University, Seoul 08826, Republic of Korea}

\author[0000-0003-4625-229X]{Woong-Tae Kim}
\affiliation{Department of Physics $\&$ Astronomy, Seoul National University, Seoul 08826, Republic of Korea}
\affiliation{SNU Astronomy Research Center, Seoul National University, Seoul 08826, Republic of Korea}

\begin{abstract}
While bars are common in disk galaxies, their formation conditions are not well understood. We use $N$-body simulations to study bar formation and evolution in isolated galaxies consisting of a stellar disk, a classical bulge, and a dark halo. We consider 24 galaxy models that are similar to the Milky Way but differ in the mass and compactness of the classical bulge and halo concentration. We find that the bar formation requires $(\QTmin/1.2)^2 + (\text{CMC}/0.05)^2\lesssim 1$, where $\QTmin$ and CMC refers to  the minimum value of the Toomre stability parameter and the central mass concentration, respectively. Bars tend to be stronger, longer, and rotate slower in galaxies with a less massive and less compact bulge and halo. All bars formed in our models correspond to slow bars. A model with the bulge mass of $\sim10$--$20$\% of the disk under a concentrated halo produces a bar similar to the Milky-Way bar. We discuss our findings in relation to other bar formation criteria suggested by previous studies.
\end{abstract}



\section{Introduction} \label{sec:intro}

Bars are common in the universe. More than $\sim60\%$ of disk galaxies in optical and near-infrared images are known to possess a weak or strong bar in the local universe \citep{de63,sw93,kna00, whyte02,lauri04,marijogee07,menendez07,agu09,mendez12,buta15,diaz16,diaz19}. The fraction of barred disk galaxies decreases with redshift \citep{sheth08,mel14}, with a tendency that more massive galaxies are more likely barred. Bar formation appears inhibited in dispersion-dominated galaxies and in halo-dominated galaxies at low redshift \citep{sheth12}. These indicate that the bar formation occurs preferentially in a late secular phase of galaxy formation when the disks become dynamically cold  \citep{kra12}.

Theoretically, bar formation is due to gravitational instability of a rotationally-supported stellar disk \citep{toomre64}: non-axisymmetric perturbations grow via swing amplification and initially circular stellar orbits are deformed to elongated $x_1$ orbits that form and support a bar (see, e.g., \citealt{sell14}). A number of simulations have shown that the presence of a dark halo affects the bar formation and evolution \citep{onp73,hohl76,deba00,vnk03,holley05,wnk07}. While the gravity of a halo tends to suppress the bar formation by reducing the relative strength of the disk's self-gravity in equilibrium \citep{onp73}, angular momentum exchange between a bar and a live halo allows the former to grow longer and stronger \citep{ath02}. 
Also, the halo parameters such as the axial ratio \citep{ath02,ath13} and spin  \citep{col18,col19,kat22}
lead to considerable changes in the bar evolution.

In addition to a halo, a classical bulge can also strongly affect the bar formation and evolution. Classical bulges are produced as a result of major/minor mergers during galaxy formation \citep{kauff93,bau96,hop09,naab14,bou07,hop10}. Unlike halos, classical bulges are highly centrally concentrated and can thus stabilize the inner regions of disks, without affecting the outer regions much. Early studies found that a strong bulge suppresses swing amplification by interrupting a feedback loop that transforms propagating trailing waves to leading ones \citep{sell80,toomre81,bnt08}, inhibiting bar formation (e.g., \citealt{se18,kd18}). Also, a live bulge can make a bar longer and stronger by removing angular momentum from the latter, just like a live halo \citep{sell80}.  A bar that forms can increase a central mass, for example, by driving gas inflows \citep[e.g.,][]{ath92,but96,kim12}, which in turns weakens or destroys the bar by disturbing bar-supporting $x_1$ orbits \citep[e.g.,][]{pfe90,has93,nor96,she04,bou05,ath13}.

While many numerical studies mentioned above are useful to understand the effects of a bulge and a halo on the bar formation and evolution, the quantitative conditions for bar-forming instability have still been under debate. Using numerical models with a fixed halo and a disk with surface density $\Sigma_d\propto R^{-1}$, \citet{onp73} suggested that bar formation requires 
\begin{equation}\label{e:tOP}
    t_\text{OP} \equiv T/|W| > 0.14,
\end{equation}
where $T$ and $W$ stand for the total rotational and gravitational potential energies of a galaxy, respectively. 
Using two-dimensional (2D) models with a fixed halo and an exponential disk, \cite{efsta82} showed that a bar forms in galaxies with  
\begin{equation}\label{e:eELN}
    \epsilon_\text{ELN}  \equiv \frac{V_{\rm max}}{(GM_{d}/R_{d})^{1/2}} < 1.1,
\end{equation}
where $V_{\rm max}$, $M_{d}$, and $R_d$ refer to the maximum rotational velocity, mass, and scale radius of the disk, respectively. 

It is not until recent years that galaxy models for bar formation treat all three components (disk, bulge, and halo) as being live \citep{poly16,salo17,se18,fujii18,kd18,kd19,kd20}. In particular, \citet{kd18} used self-consistent $N$-body simulations with differing bulge masses, and showed that bar formation requires that the ratio of the bulge to total radial force initially satisfies
\begin{equation}\label{e:FKD}
   \mathcal{F}_\text{KD} \equiv \frac{GM_b}{R_dV_\text{tot}^2} < 0.35,
\end{equation}
where $M_b$ is the bulge mass and $V_\text{tot}$ is the total rotational velocity at $R=R_d$. Using three-component galaxy models with differing disk and bulge densities, 
\cite{se18} argued that their models evolve to barred galaxies provided \begin{equation}\label{e:SE}
   \mathcal{D}_\text{SE} \equiv \frac{\left<\rho_b\right>}{\left<\rho_d\right>} < \frac{1}{\sqrt{10}},
\end{equation}
where $\left<\rho_b\right>$ and $\left<\rho_d\right>$ are the mean densities of the bulge and disk, respectively, within the half-mass radius of the bulge.

The several different conditions given above imply that there has not been consensus regarding the quantitative criterion for bar formation. A part of the reason for the discrepancies in the proposed conditions may be that some models considered a fixed (rather than live) halo, and that some authors explored parameter space by fixing either bulge or halo parameters. Also, it is questionable whether the effects of the complicated physical processes (swing amplification and feedback loop) involved in the bar formation can be encapsulated by the single parameters given above. In this paper, we revisit the issue of bar formation by varying both bulge and halo parameters altogether.  Our models will be useful to clarify what conditions are necessary to produce a bar when the mass and compactness of the bulge and halo vary.
We will show that the two key elements that govern the bar formation  are the minimum value of the Toomre stability parameter $\QTmin$ and the central mass concentration (CMC), defined as the total galaxy mass inside the central $0.1\kpc$ relative to the total disk mass: bars form more easily in galaxies with smaller $\QTmin$ and CMC.    
We also measure the strength, length, and pattern speed of the bars that form in our simulations and explore their dependence on the halo and bulge parameters. 

This paper is organized as follows. In \autoref{sec:modelnmethod}, we describe our galaxy models and numerical methods we employ. In \autoref{sec:results}, we present temporal changes of the bar properties such as bar strength, pattern speed, length, and angular momentum transfer from a disk to halo and bulge. In \autoref{sec:discussion}, we compare our numerical results with the previous bar formation conditions mentioned above, and propose the new conditions in terms of $\QTmin$ and CMC. We also use our numerical model to constrain the classical bulge of the Milky Way. Finally, we conclude our work in \autoref{sec:summary}.

\begin{deluxetable*}{ccccccccccccc}
\tablecaption{Model parameters and various dimensionless quantities of the initial galaxy models   \label{tbl:model}}
\tablenum{1}
\tablehead{
\colhead{Model} & 
\colhead{$a_{h}$} & 
\colhead{$M_b/M_d$} & 
\colhead{$a_{b}$} & 
\colhead{$\QTmin$} & 
\colhead{CMC}&
\colhead{$t_\text{OP}$}&
\colhead{$\epsilon_\text{ELN}$} & 
\colhead{$\mathcal{F}_\text{KD}$} &
\colhead{$\mathcal{D}_\text{SE}$} \\
\colhead{} & 
\colhead{(kpc)} & 
\colhead{} & 
\colhead{(kpc)} & 
\colhead{} &
\colhead{} &
\colhead{} & 
\colhead{} & 
\colhead{} & 
\colhead{}  \\
\colhead{(1)} & 
\colhead{(2)} & 
\colhead{(3)} & 
\colhead{(4)} & 
\colhead{(5)} &
\colhead{(6)} & 
\colhead{(7)} & 
\colhead{(8)} & 
\colhead{(9)} &
\colhead{(10)}}
\startdata
\texttt{C00}   & 30         & 0.0     & 0.4  & 1.062  &$0.04\times10^{-2}$    &0.449 & 0.89   & 0.0  & 0.0\\
\texttt{C05}   & 30         & 0.05  & 0.4  & 1.079    &$0.24\times10^{-2}$   &0.449& 0.90   & 0.10  & 0.67\\
\texttt{C10}   & 30         & 0.1   & 0.4  & 1.085    &$0.44\times10^{-2}$   &0.447& 0.91   & 0.19 & 1.30 \\
\texttt{C20}   & 30         & 0.2   & 0.4  & 1.110    &$0.84\times10^{-2}$   &0.446& 0.92   & 0.33 & 2.60 \\
\texttt{C30}   & 30         & 0.3   & 0.4  & 1.140    &$1.22\times10^{-2}$   &0.445& 0.94   & 0.44  & 3.90\\
\texttt{C40}   & 30         & 0.4   & 0.4  & 1.164    &$1.62\times10^{-2}$   &0.444& 0.95   & 0.52  & 5.20\\
\hline
\texttt{L00}   & 40         & 0.0     & 0.4  & 0.954  &$0.04\times10^{-2}$     &0.442& 0.80   & 0.0  & 0.0 \\
\texttt{L05}   & 40         & 0.05  & 0.4  & 0.961    &$0.24\times10^{-2}$   &0.441& 0.81   & 0.12  & 0.64 \\
\texttt{L10}   & 40         & 0.1   & 0.4  & 0.975    &$0.44\times10^{-2}$   &0.440& 0.82   & 0.22  & 1.30\\
\texttt{L20}   & 40         & 0.2   & 0.4  & 1.000    &$0.84\times10^{-2}$   &0.439& 0.84   & 0.37 & 2.61 \\
\texttt{L30}   & 40         & 0.3   & 0.4  & 1.023    &$1.24\times10^{-2}$   &0.438& 0.86   & 0.49  & 3.91\\
\texttt{L40}   & 40         & 0.4   & 0.4  & 1.046    &$1.64\times10^{-2}$   &0.437& 0.89   & 0.58  & 5.19\\
\texttt{L50}   & 40		&0.5	   &0.4   &1.056  &$2.06\times10^{-2}$  &  0.436 &0.99& 0.64&6.05 \\
\hline
\texttt{C05c}  & 30         & 0.05  & 0.2  & 1.082    &$0.58\times10^{-2}$   &0.447& 0.90   & 0.10  & 3.01\\
\texttt{C10c}  & 30         & 0.1   & 0.2  & 1.086    &$1.16\times10^{-2}$   &0.446& 0.91   & 0.18  & 5.76\\
\texttt{C20c}  & 30         & 0.2   & 0.2  & 1.106    &$2.26\times10^{-2}$   &0.442& 0.92   & 0.32  & 11.30\\
\texttt{C30c}  & 30         & 0.3   & 0.2  &  1.127   &$3.34\times10^{-2}$   &0.439&1.07    & 0.42  &16.80 \\
\texttt{C40c}  & 30         & 0.4   & 0.2  &   1.158  &$4.40\times10^{-2}$   &0.437&1.23    &0.50   & 22.21\\
\hline
\texttt{L05c}  & 40         & 0.05  & 0.2  & 0.961    &$0.58\times10^{-2}$   &0.439& 0.81   & 0.12  & 2.89\\
\texttt{L10c}  & 40         & 0.1   & 0.2  & 0.972    &$1.16\times10^{-2}$   &0.438& 0.82   & 0.21 & 5.72 \\
\texttt{L20c}  & 40         & 0.2   & 0.2  & 0.999    &$2.26\times10^{-2}$   &0.435& 0.88   & 0.36  & 11.28\\
\texttt{L30c}  & 40         & 0.3   & 0.2  & 1.024    &$3.38\times10^{-2}$   &0.432& 1.07   & 0.46 & 16.84 \\
\texttt{L40c}  & 40         & 0.4   & 0.2  & 1.049    &$4.48\times10^{-2}$   &0.430& 1.23   & 0.54  & 22.06\\
\texttt{L50c}  & 40         & 0.5   & 0.2  & 1.049  &$5.62\times10^{-2}$    & 0.428     &1.38    &0.59   &25.53 
\enddata
\end{deluxetable*}
\section{Galaxy Model and method} \label{sec:modelnmethod}
\subsection{Galaxy Models} \label{subsec:models}

To study the effects of spheroidal components on the bar formation and evolution in disk galaxies, we consider Milky Way-like, isolated galaxies. Our galaxy models are three dimensional (3D), consisting of a dark matter halo, a classical bulge, a stellar disk, and a central supermassive black hole.

For the stellar disk, we adopt the exponential-secant hyperbolic density distribution
\begin{equation}
\rho_d(R,z) = \frac{M_d}{4\pi z_dR_d^2} \exp\left( -\frac{R}{R_d}\right)
{\rm sech}^2\left( \frac{z}{z_d}\right),
\end{equation}
where $R$ is the cylindrical radius, $R_d$ is the disk scale radius, $z_d$ is the disk scale height, and $M_d$ is the total disk mass.
We fix $R_d=3\kpc$, $z_d=0.3\kpc$, and $M_d=5\times 10^{10}\Msun$, similar to the Milky Way (\citealt{bg16}, \citealt{helmi20}). Initially, we set the velocity anisotropy to $f_R\equiv\sigma_R^2/\sigma_z^2=2.0$,
where $\sigma_R$ and $\sigma_z$ refer to the velocity dispersions of the  disk particles in  the radial and vertical directions, respectively.
We note that $f_R$ increases with time as the disk evolves to form a bar, becoming similar to the observed value of $f_R \sim 4$ near the solar neighborhood (e.g., \citealt{Sha14,Gui15,Katz18}). 

\begin{figure}[t]
\centering
\epsscale{1.0} \plotone{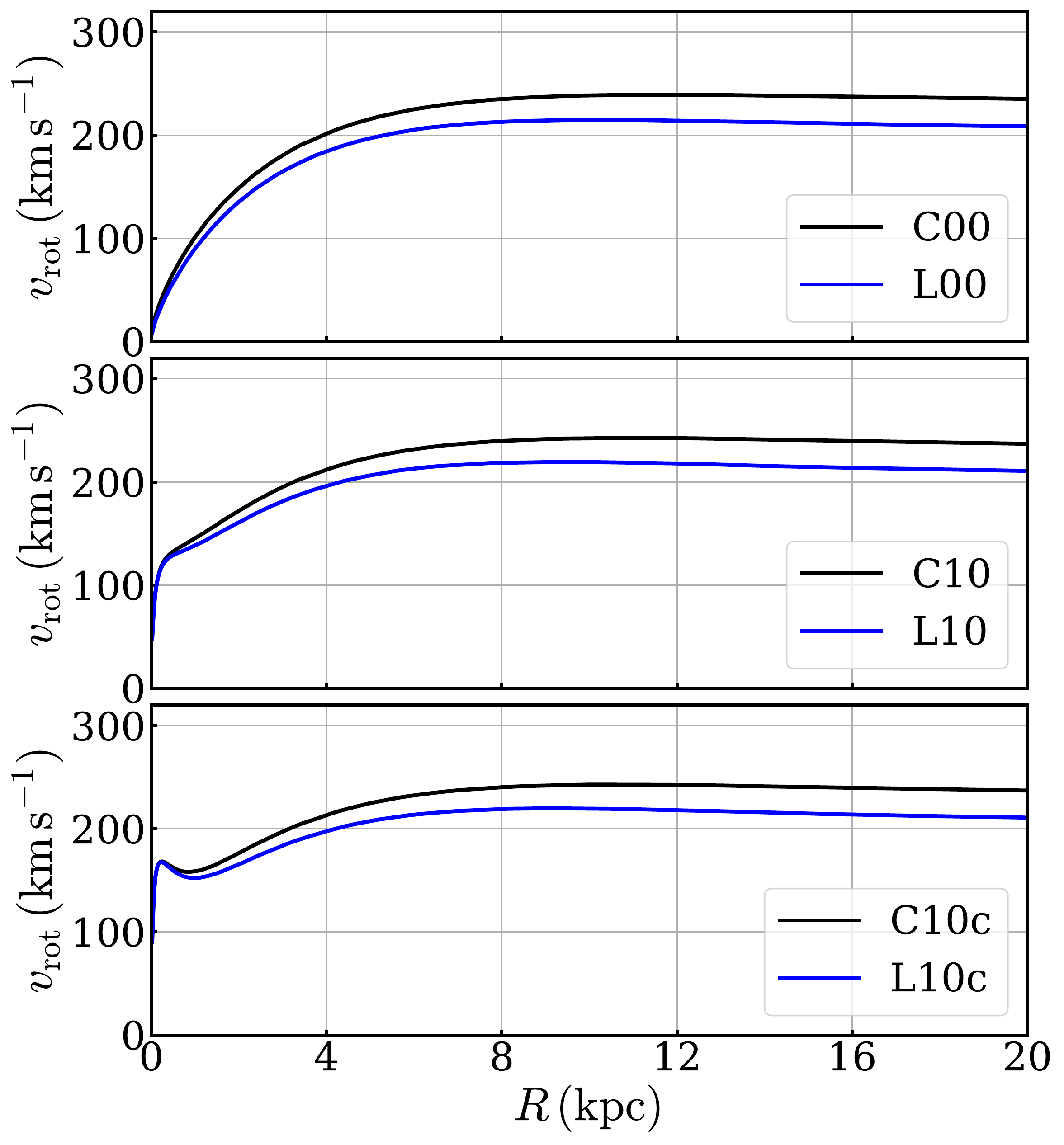}
\caption{Radial distributions of the total rotational velocity $v_\textrm{rot}$ for models \texttt{C00}, \texttt{L00} (top), \texttt{C10}, \texttt{L10} (middle), and \texttt{C10c}, \texttt{L10c} (bottom). A more massive and compact bulge increase $v_\textrm{rot}$ at small $R$. Models in the \texttt{C} series have higher $v_\textrm{rot}$, by $\sim20\kms$ on average, than in the \texttt{L} series. 
\label{fig:vrot_LC}}
\end{figure}

For both halo and classical bulge, we take the \cite{hernquist90}  profile
\begin{equation}
\rho(r) = \frac{M}{2\pi}\frac{a}{r(r+a)^3},
\end{equation}
where $r=(R^2+z^2)^{1/2}$ is the spherical radius, and $M$ and $a$ denote the mass and the scale radius of each component, respectively.
For the halo, we fix its mass to $M_{h}=1.35\times 10^{12}\Msun=26M_d$ and consider two scale radii: a centrally concentrated halo with $a_h=30\kpc$ and a less concentrated halo with $a_h=40\kpc$, which we term \texttt{C} and \texttt{L} series, respectively.  
For the bulge, we vary both mass $M_b=(0$--$0.5)M_d$ and scale radius $a_b$ between $0.2$ to $0.4\kpc$. 
We place a supermassive black hole with mass $M_\mathrm{BH}=4 \times 10^{6}\Msun$ at the galaxy center.

\autoref{tbl:model} lists the names and initial parameters of all models. Column (1) gives the model names. Column (2) gives the scale radius of the halo, while Columns (3) and (4) list  the bulge mass relative to the disk mass and the bulge scale radius, respectively.  The prefix \texttt{C} and \texttt{L} in the model names stand for the centrally concentrated and less concentrated halo, respectively. The number after the prefix denotes the bulge mass relative to the total disk mass. The postfix \texttt{c} implies a compact bulge: the models with and without the postfix have $a_{b} = 0.2\kpc$ and $0.4\kpc$, respectively. 
For example, model \texttt{L20c} has a less-concentrated halo with $a_h=40\kpc$ and a compact bulge with $M_b=0.2M_d$ and $a_{b} = 0.2 \kpc$. Column (5) lists the minimum value of the Toomre stability parameter. Column (6) lists the CMC. Columns (7)--(10) give the values for the quantities defined in \crefrange{e:tOP}{e:SE}. We take model {\tt C10} as our fiducial model. 

\begin{figure}
    \centering
    \plotone{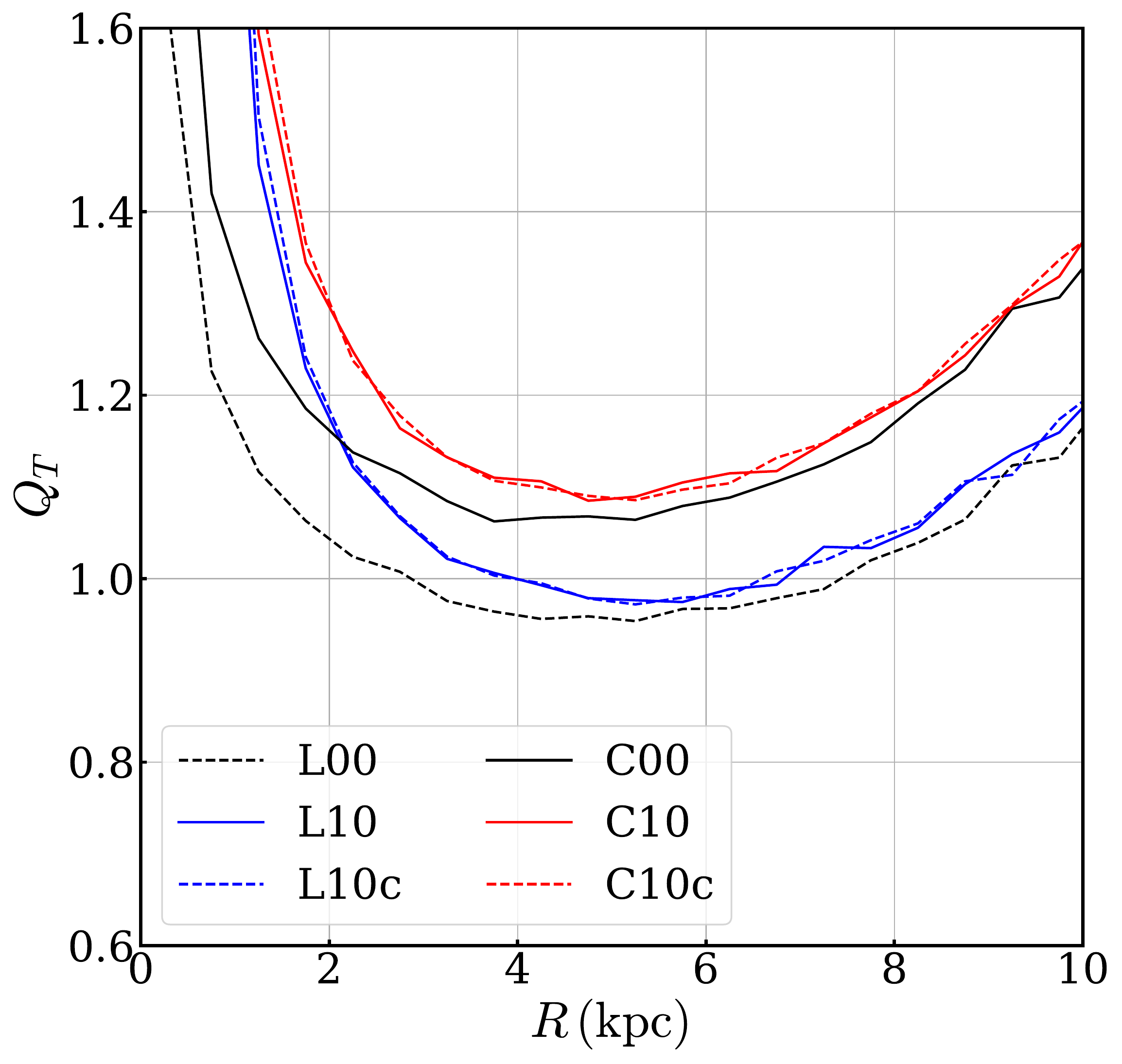}
    \caption{Radial profiles of $\QT$ for models with $M_b/M_d=0, 0.1$. The minimum value $\QTmin$ occurs at $R\sim4$--$6\kpc$, which tends to be larger for a galaxy with a more concentrated halo and/or more massive bulge.
    }
    \label{fig:toomreQ}
\end{figure}

\Cref{fig:vrot_LC} plots the radial distributions  of the total rotational velocity $v_\textrm{rot}$ for selected models. The black and blue lines correspond to the models in the \texttt{C} and \texttt{L} series, respectively. 
It is apparent that increasing the bulge mass enhances the rotational velocity.  Models with a compact bulge have higher rotational velocity in the inner regions with $R\lesssim a_b$. At $R\lesssim 20\kpc$, models in the \texttt{C} series have larger $v_\textrm{rot}$, by $\sim 20\kms$ on average, than the \texttt{L} series counterparts.

The gravitational susceptibility of a disk can be measured by 
the \cite{toomre66} stability parameter  
\begin{equation}
   \QT = \frac{\kappa \sigma_R}{3.36G\Sigma_d},
\end{equation}
where $\kappa$ is the epicycle frequency and $\Sigma_d$ is the disk surface density. \cref{fig:toomreQ} plots the radial distributions of $\QT$ for models with $M_b/M_d=0$ and $0.1$. Overall, $\QT$ is large at both small $R$ (due to increase in $\kappa$) and large $R$ (due to decrease in $\Sigma_d$) and attains a minimum value $\QTmin$ at $R\sim4$--$6\kpc$. 
Our galaxy models have $\QTmin$ in the range between 0.95 and 1.16 (\autoref{tbl:model}): $\QTmin$ tends to be larger for a galaxy with a centrally concentrated halo and/or more massive bulge, while it is almost independent of the bulge compactness.

\subsection{Numerical Method} \label{subsec:method}

To construct the initial galaxy models, we make use of the GALIC code
\citep{yu14} which solves the collisionless Boltzmann equations to find a desired equilibrium state by optimizing the velocities of individual particles.
We distribute $N_d = 1.0 \times 10^6$, and $N_b = 5 \times 10^4$--$5\times 10^5$, and $N_h = 2.6 \times 10^7$ particles for the disk, bulge, and halo,  respectively. We set the mass of each particle to $m=5\times10^4\Msun$, which are equal for all three components. 

\begin{figure*}[ht]
 \centering
\epsscale{1.0} \plotone{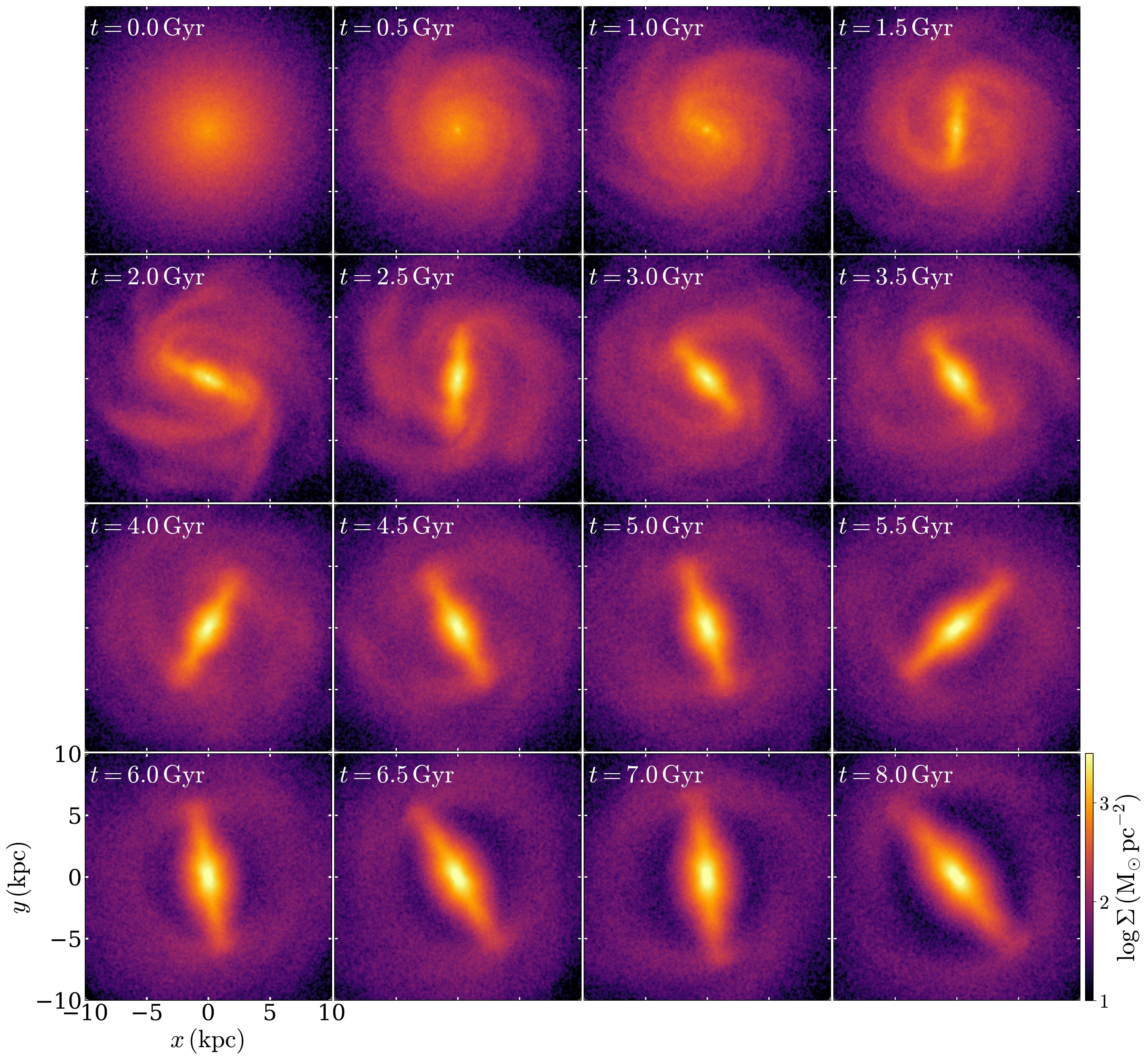}
 \caption{Snapshots of the disk surface density in model  \texttt{C10}. \label{fig:snapC10}}
 \end{figure*}

We evolve our galaxy models by using a public version of the Gadget-4 code \citep{springel21}. This version has improved force accuracy, time-stepping, computational efficiency, and parallel scalability from Gadget-3. It offers the Fast Multipole Method in which the tree is accelerated by multipole expansion not only at the source side but also at the sink side. 
For our galaxy models, we find the multipole expansion to order $p=4$ is fastest. In addition, hierarchical time-integration scheme can effectively reduce the computation time by constructing a tree only for the set of particles involved in the current force calculation.
We take the force accuracy parameter $\alpha=3\times10^{-4}$ which conserves the total angular momentum within $\sim0.1$ percent (see below). The softening parameters for dark halo, stellar disk, and bulge particles are set to $0.05\kpc$, $0.01\kpc$, and $0.01\kpc$, respectively.

\section{Results} \label{sec:results}

In this section, we present evolution of our models with a focus on the temporal changes in the strength, pattern speed, and size of the bars that form. The bar formation conditions will be discussed  in  \autoref{sec:discussion}. 

\subsection{Bar Formation and Strength} \label{subsec:bar}

 \begin{figure*}[ht]
 \centering
\epsscale{0.90} \plotone{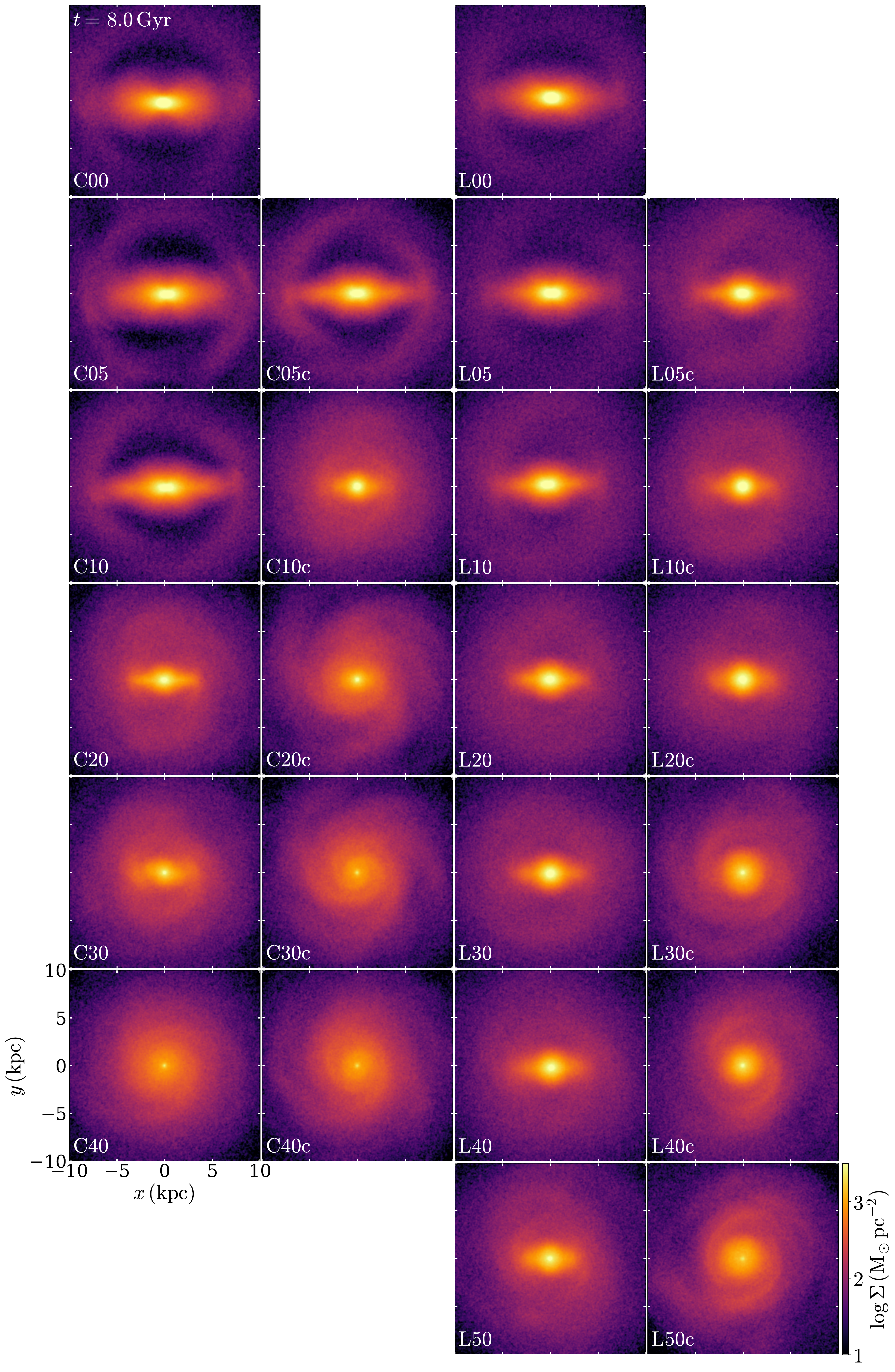}
 \caption{Snapshots of the disk surface density at $t=8.0\Gyr$ in all the models.  Each image is rotated such that the semi-major axis of a bar (or an oval) is aligned parallel to the $x$-axis.
 \label{fig:snapall}}
 \end{figure*}

\autoref{fig:snapC10} plots snapshots of the disk surface density for our fiducial model \texttt{C10}.
\autoref{fig:snapall} plots the snapshots of all the models at {$t= 8.0\Gyr$}.
In model \texttt{C10} with $\QTmin=1.09$, non-axisymmetric perturbations inherent in the particle distributions grow as they swing from leading to trailing configurations (e.g., \citealt{bnt08,kim07,kwak17,seo19}), forming spiral arms at $t=0.5\Gyr$. Since the inner Lindblad resonance (ILR) is weak in this model, trailing spiral waves propagate toward the galaxy center to become leading waves at the opposite side, which can grow further: successive swing amplifications combined with multiple feedback loops eventually lead to a bar at $t\gtrsim1.5\Gyr$.  If $\QTmin$ is quite small, as in models \texttt{C00} and \texttt{L00}, the swing amplification is so virulent that the inner parts of the spiral arms are rapidly organized into a bar in less than $\sim1\Gyr$. In contrast, if $\QTmin$ is large or the ILR is too strong, as in models \texttt{C40}, \texttt{C20c}, and \texttt{L30c},  the feedback loop is blocked and the disks produce only spirals, sometimes with an oval. 

\begin{figure} 
\centering
\plotone{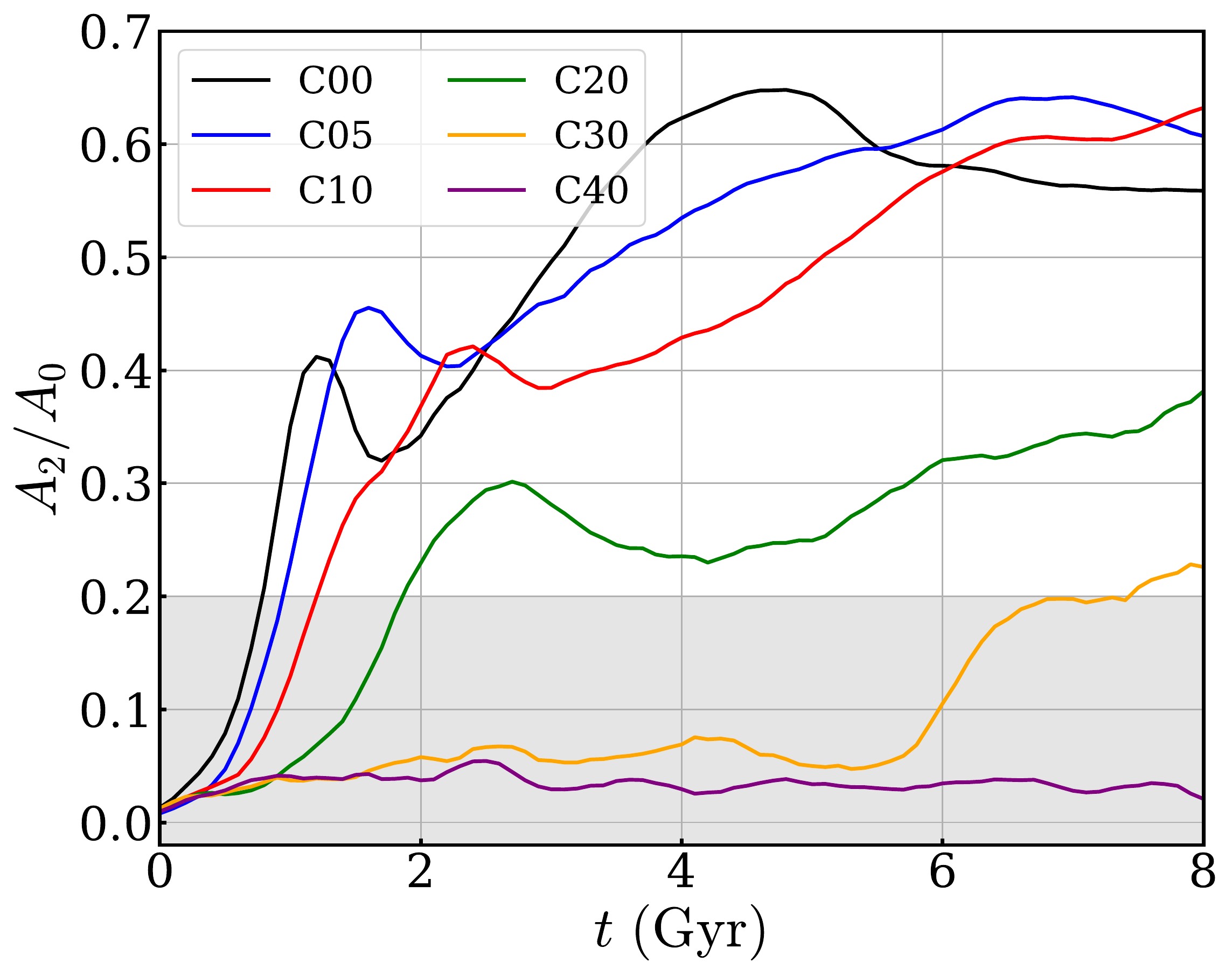}
\plotone{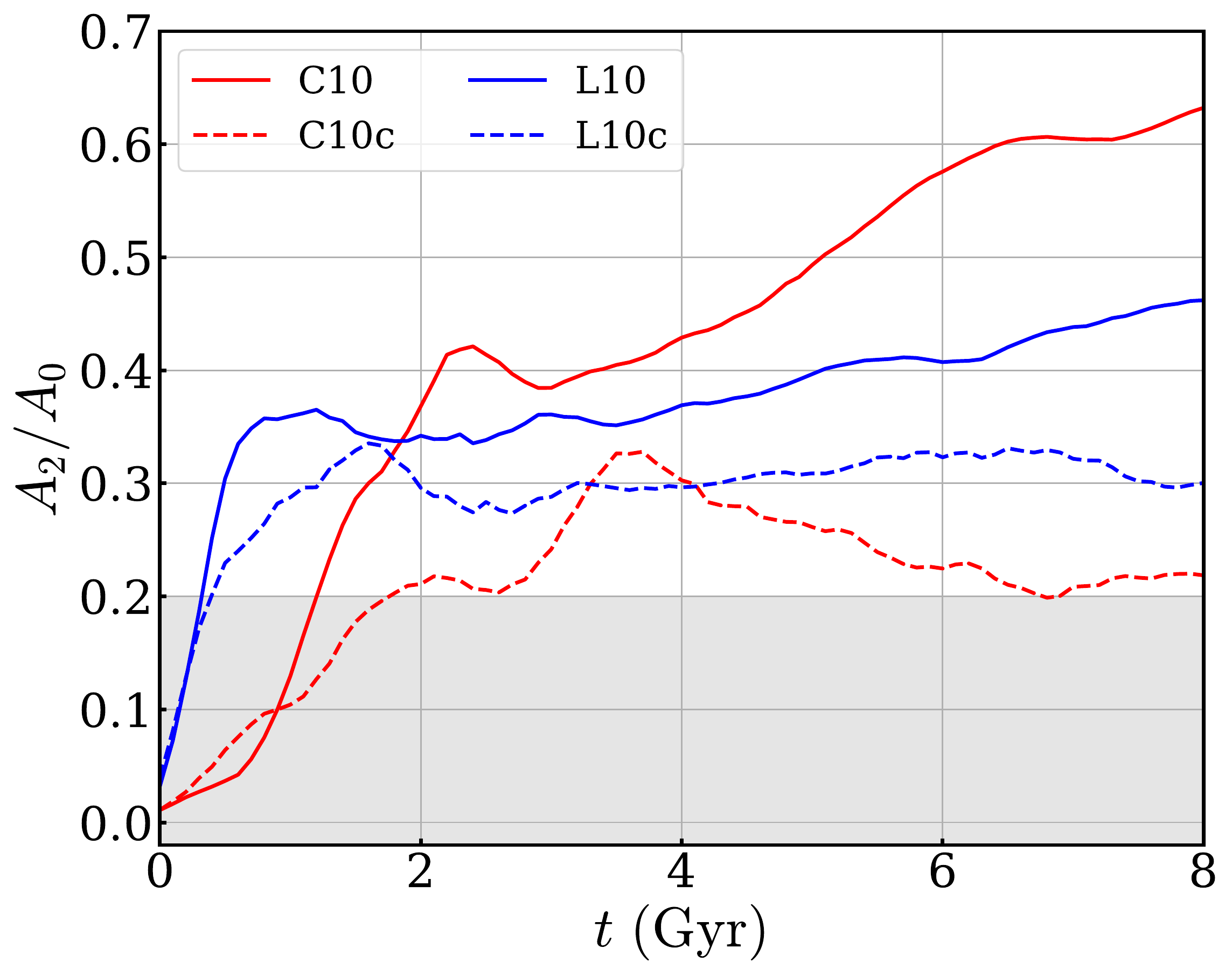}
\caption{Temporal changes of the bar strength $A_2/A_0$ 
for models with less compact bulges in the \texttt{C} series (top) and for models with $M_b/M_d=0.1$ (bottom). Features with $A_2/A_0\leq 0.2$ are regarded as ovals or spirals.\label{fig:A2_LC}}
\end{figure}

To quantify the bar strength, we first consider an annular region of the disk centered at radius $R$ with width $\Delta R=1\kpc$ and calculate the amplitudes of the $m=2$ Fourier modes as
\begin{subequations}
\begin{align}
    a_2(R) &=  \sum_{i} m_i \cos{(2\theta_i)}, \\ b_2(R) &=  \sum_{i} m_i \sin{(2\theta_i)},
\end{align}
\end{subequations}
where $\theta_i$ and $m_i$ are the azimuthal angle and mass of the $i$-th disk particle in the annulus, respectively. We then define the bar strength as
\begin{equation}\label{e:barstr}
   \frac{A_2}{A_0} = \textrm{max} \left(\frac{\sqrt{a_2^2+b_2^2}}{\Sigma_{i}m_i}\right).
\end{equation}
Note that $A_2/A_0$ measures the strength of $m=2$ spirals when a bar is absent or weak. For spirals,  the position angle 
\begin{equation}\label{e:posang}
 \psi(R)\equiv \frac{1}{2} \tan^{-1}\left(\frac{b_2}{a_2}\right)
\end{equation} 
systematically varies with $R$, while $\psi(R)$ remains almost constant for a bar.

Following \cite{algorry17}, we regard galaxies with $A_2/A_0 \geq 0.2$ and relatively constant $\psi(R)$ as being barred: features with $A_2/A_0 < 0.2$ are considered as ovals if $\psi(R)$ is constant or spirals if $\psi(R)$ changes with $R$.
\autoref{fig:A2_LC} plots temporal evolution of $A_2/A_0$ for models with less compact bulges in the \texttt{C} series (upper panel) and for models with $M_b/M_d=0.1$ (lower panel). The evolution of $A_2/A_0$ is dominated by spirals at early time ($\lesssim1$--$2\Gyr$) and then by a bar. Although the spirals are strong in the outer regions, they can affect the inner disk where a bar exists before it fully grows (see the $t\leq 3\Gyr$ panels in \autoref{fig:snapC10}). The spirals and bar rotate about the galaxy center at different pattern speeds. When the spirals and bar become in phase, $A_2/A_0$ achieves its peak value temporarily (at $t=2.3\Gyr$ for model \texttt{C10}). Subsequently, $A_2/A_0$ decreases as they become out of phase, although it increases again as the bar grows further and dominates the inner disk. The presence of a more massive bulge makes the bar forms later and weaker. The bar formation is completely suppressed in models with $M_b/M_d\geq 0.4$ in the \texttt{C} series.

The compactness of halo and bulge has a significant effect on the bar formation. In the \texttt{L} series with a less concentrated halo, disks are unstable to form a bar even when the bulge mass amounts to $\sim50\%$ of the disk mass. In contrast, disks in the \texttt{C} series with a compact halo do not produce a bar when $M_b/M_d \gtrsim 0.35$.
Similarly, a more compact bulge tends to suppress the bar formation. 
For example, the maximum bulge mass for bar formation is decreased to 20\% and 10\% in the \texttt{L} and \texttt{C} series, respectively, when the bulge is compact. 
Our result that a bar does not form in galaxies with a very massive and compact bulge is qualitatively consistent with previous studies (e.g., \citealt{kd18,se18}).

\subsection{Buckling Instability} \label{subsec:buckling}

 \begin{figure}[t]
 \centering
 \plotone{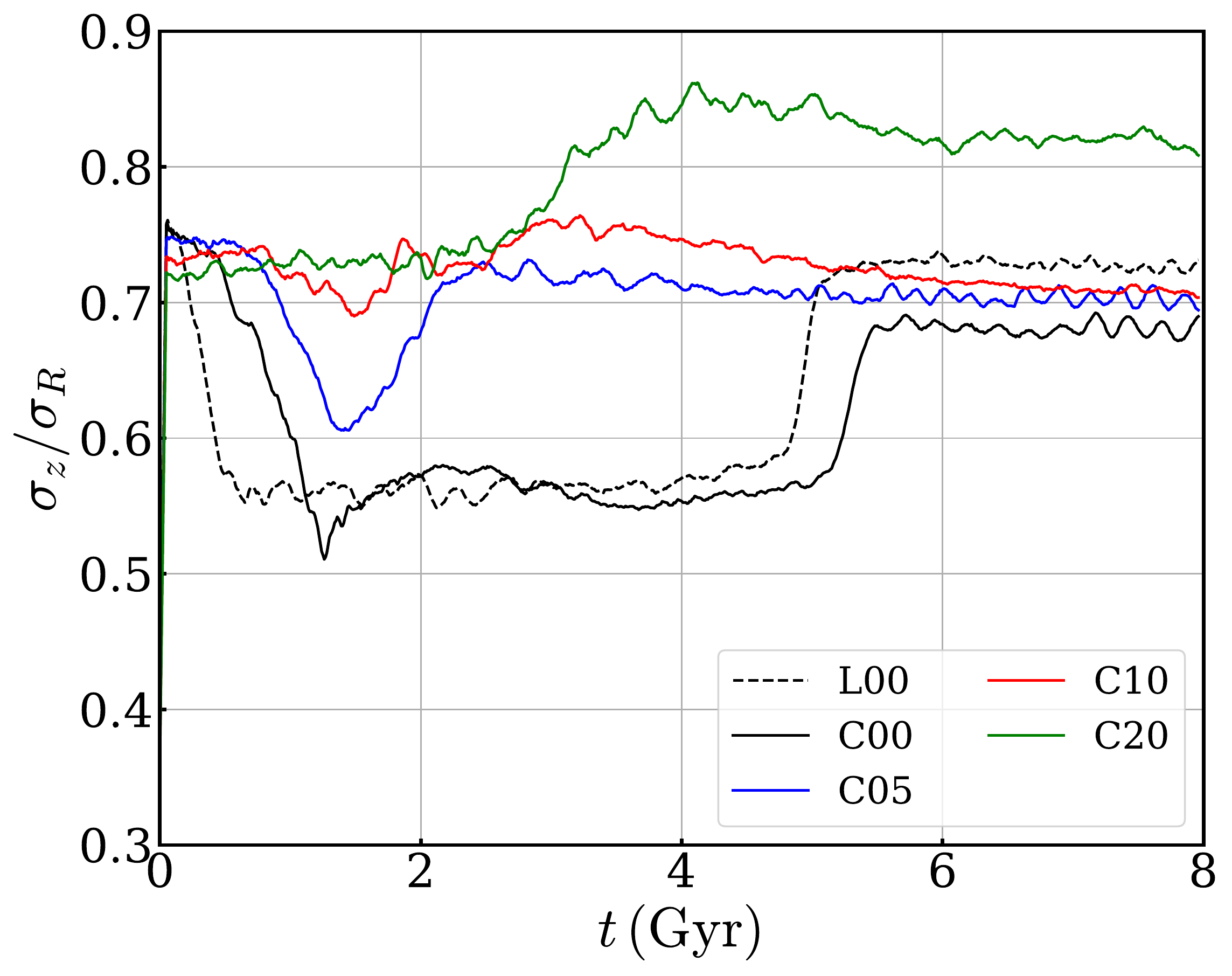}
 \caption{Time evolution of the ratio, $\sigma_z/\sigma_R$, of the velocity dispersions at $R=1\kpc$ for models with $M_b/M_d\le 0.2$ in the \texttt{C} series together with model \texttt{L00}. A rapid increase of $\sigma_z/\sigma_R$ at $t\sim5\Gyr$ in models \texttt{C00} and \texttt{L00} is due to buckling instability.
 \label{fig:Sigma}}
  \end{figure}

\begin{figure}[t]
\centering
\plotone{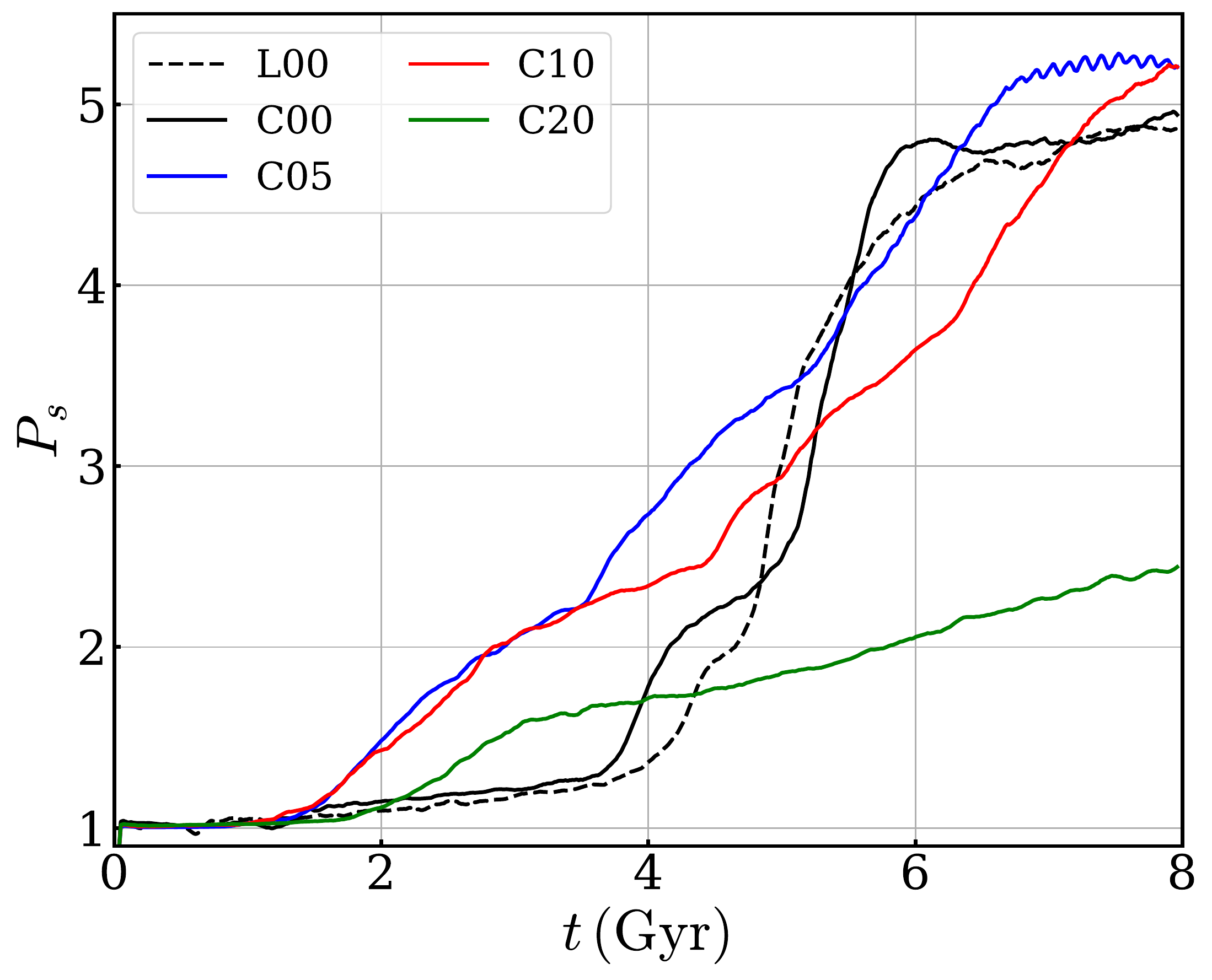}
 \caption{Time evolution of the B/P strength, $P_s$, defined in  \cref{eq:Ps}, for models shown in \autoref{fig:Sigma}. In most models, $P_s$ increases relatively slowly, while it increases rapidly at $t\sim5\Gyr$ in models \texttt{C00} and \texttt{L00}, corresponding to buckling instability. \label{fig:Ps}}
 \end{figure}

 \begin{figure}[t]
 \centering
\plotone{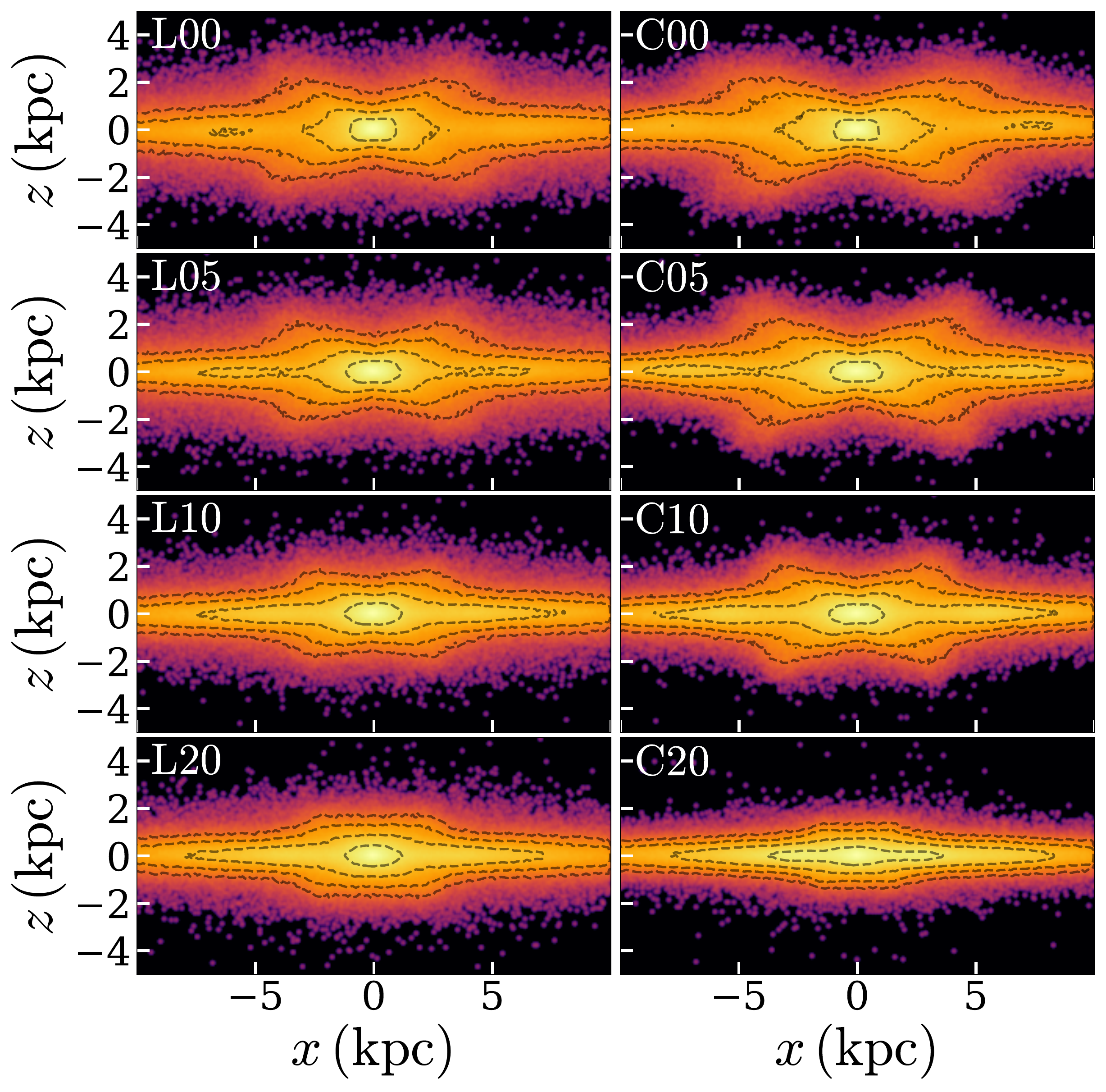}
 \caption{Contours of logarithm of the projected disk densities at $t=6.0\Gyr$ in models with $M_b/M_d\leq0.2$. The $x$- and $z$-axes correspond to the bar semi-major axis and the vertical direction, respectively. The dotted contours denote $\Sigma=10^{9.5}, 10^{9.0}, 10^{8.5}, 10^{8.0}\Msun\kpc^{-2}$  from inside to outside. \label{fig:density}}
 \end{figure}

\autoref{fig:Sigma} plots the temporal changes in the ratio, $\sigma_z/\sigma_R$, of the vertical to radial velocity dispersion of the disk particles at $R=1\kpc$ for models with $M_b/M_d\leq 0.2$ in the {\tt C} series together with model \texttt{L00}. Since a bar is supported by $x_1$ orbits elongated along the bar semi-major axis, its growth naturally involves the increase in $\sigma_R$. At the same time, the bar and spirals can excite the vertical motions of star particles, enhancing $\sigma_z$ (e.g., \citealt{qui14}). When a bar grows rapidly, as in models \texttt{C00} and \texttt{L00}, $\sigma_R$ increases faster than $\sigma_z$, resulting in a decrease in  the ratio $\sigma_z/\sigma_R$ at early time. When a bar grow slowly, in contrast, $\sigma_z/\sigma_R$ remains more or less constant.  

The increase in $\sigma_z$ leads to the disk thickening and the formation of a boxy/peanut (B/P) bulge. All the bars that form in our models evolve to B/P bulges. \autoref{fig:Ps} plots evolution of the B/P strength, defined as
\begin{equation}\label{eq:Ps}
    P_s = \text{max}\left(\frac{\tilde{|z|}}{|\widetilde{z_0}|}\right),
\end{equation}
where the tilde denotes the median and $z_0$ is the initial value \citep{ian15,frag17,seo19}, for models with $M_b/M_d\leq 0.2$.  In most models, the disk thickening occurs secularly. However, $P_s$ (as well as $\sigma_z/\sigma_R$) in  models \texttt{C00} and \texttt{L00} increases rapidly at $t\sim5\Gyr$, which is due to vertical buckling instability. 

It is well known that a bar can undergo buckling instability when $\sigma_z/\sigma_R$ is small. 
\cite{toomre66} and \cite{ara87} suggested that non-rotating thin disks are unstable to the buckling instability if $\sigma_z/\sigma_R \le 0.3$. For realistic disks with spatially varying $\sigma_z/\sigma_R$, \cite{raha91} found that the buckling instability occurs if $\sigma_z/\sigma_R \lesssim 0.25$--$0.55$ in the mid-disk regions. 
By varying the values of $\sigma_z/\sigma_R$ in $N$-body simulations, \cite{mar06} suggested that the critical value is at $\sigma_z/\sigma_R \sim0.6$ (see also \citealt{kwak17}). This is consistent with our numerical results that  models \texttt{L00} and \texttt{C00} have  $\sigma_z/\sigma_R \lesssim0.6$ before undergoing the buckling instability, as shown in \autoref{fig:Sigma}.

\autoref{fig:density} plots the edge-on views of the projected density distributions of the disks at $t=6.0 \Gyr$ for models with $M_b/M_d\leq 0.2$. The $x$- and $z$-directions correspond to the direction parallel to the bar semi-major axis and the vertical direction, respectively. The density distributions in models \texttt{L00} and \texttt{C00} with no bulge are asymmetric with respect to the $z=0$ plane, evidencing the operation of buckling instability \citep{mar06}. We note that the other models with a bulge also posses a B/P bulge which develops on a timescale longer than in models \texttt{L00} and \texttt{C00}.  This is consistent with \citet{sg20} who showed that the presence of a nuclear mass with only a small ($\sim2.5\%$) fraction of the disk mass tends to suppress the buckling instability. In models with a bulge, the disks appear to thicken as the bar particles are excited vertically by the passage through the $2:1$ vertical resonance \citep{qui14,sg20}.

\subsection{Angular Momentum and Pattern Speed} \label{subsec:omega}

\begin{figure}[t]
\centering
\epsscale{1.0}\plotone{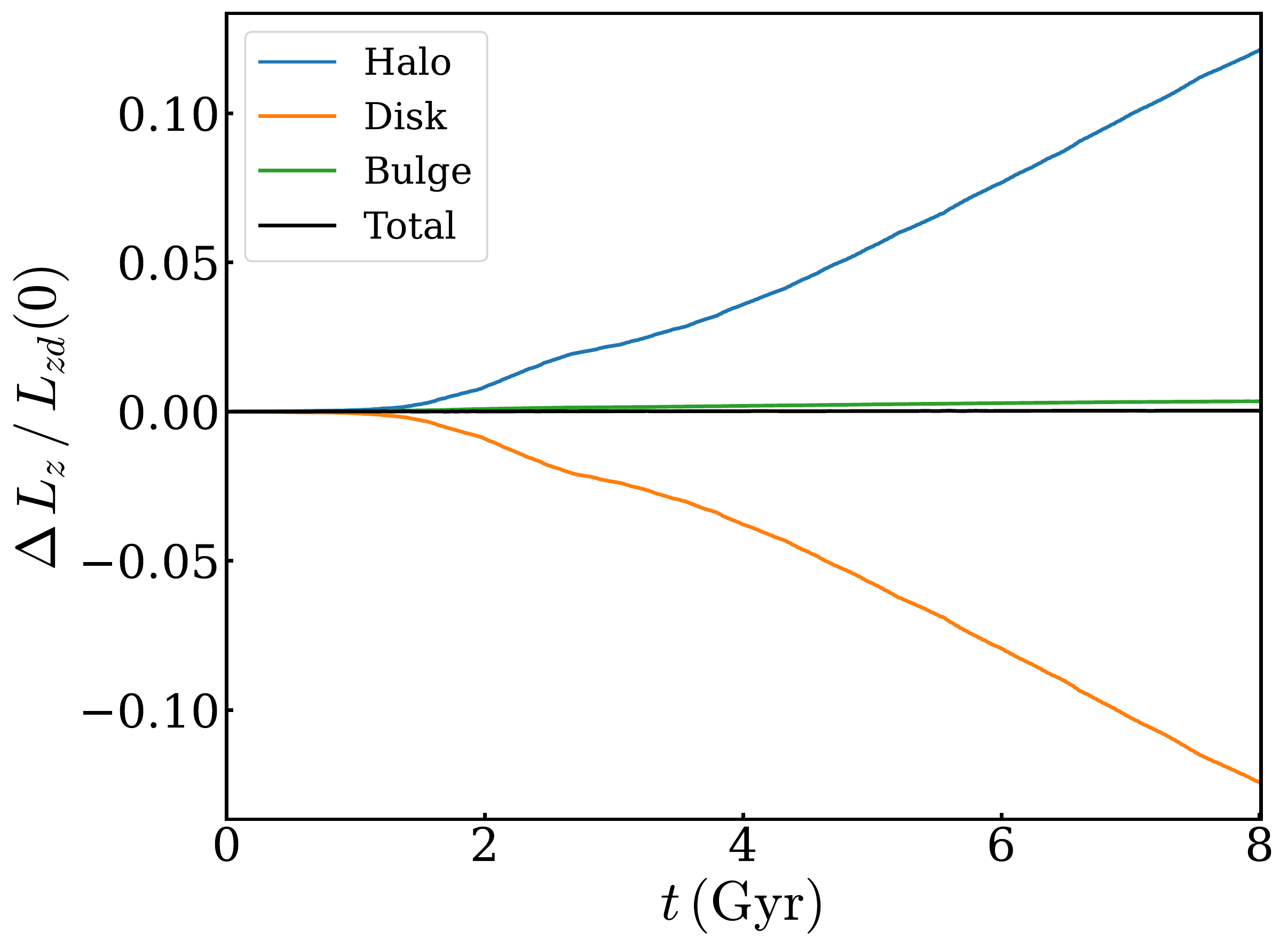}
\caption{Temporal changes of the angular momentum of the disk (orange), halo (blue), bulge (green), and the total (black) for model \texttt{C10}. All angular momenta and their changes are relative to the initial angular momentum of the disk $L_{zd}(0)$.
\label{fig:Lz}}
\end{figure}

We calculate the angular momentum of each component as 
\begin{equation}
   L_z = \sum_{i}m_i (xv_y-yv_x).
\end{equation}
\autoref{fig:Lz} plots temporal changes of $L_z$ relative to the initial disk angular momentum for a disk (orange), halo (blue), bulge (green), as well as the total (black) in model \texttt{C10}. The disk loses its angular momentum right after the bar formation, while the halo and bulge absorb it.  
Since the bulge occupies relatively a small volume in space in model \texttt{C10}, the amount of angular momentum it gains is limited to $\sim4\%$, while the halo absorbs the remaining $\sim96\%$. In model \texttt{L50} with a large bulge mass, however, the bulge absorbs about $\sim26\%$ of the angular momentum lost by the disk. The total angular momentum is conserved within $\sim 0.1\%$ in all models. 
 
\begin{figure}[t] 
\centering
\plotone{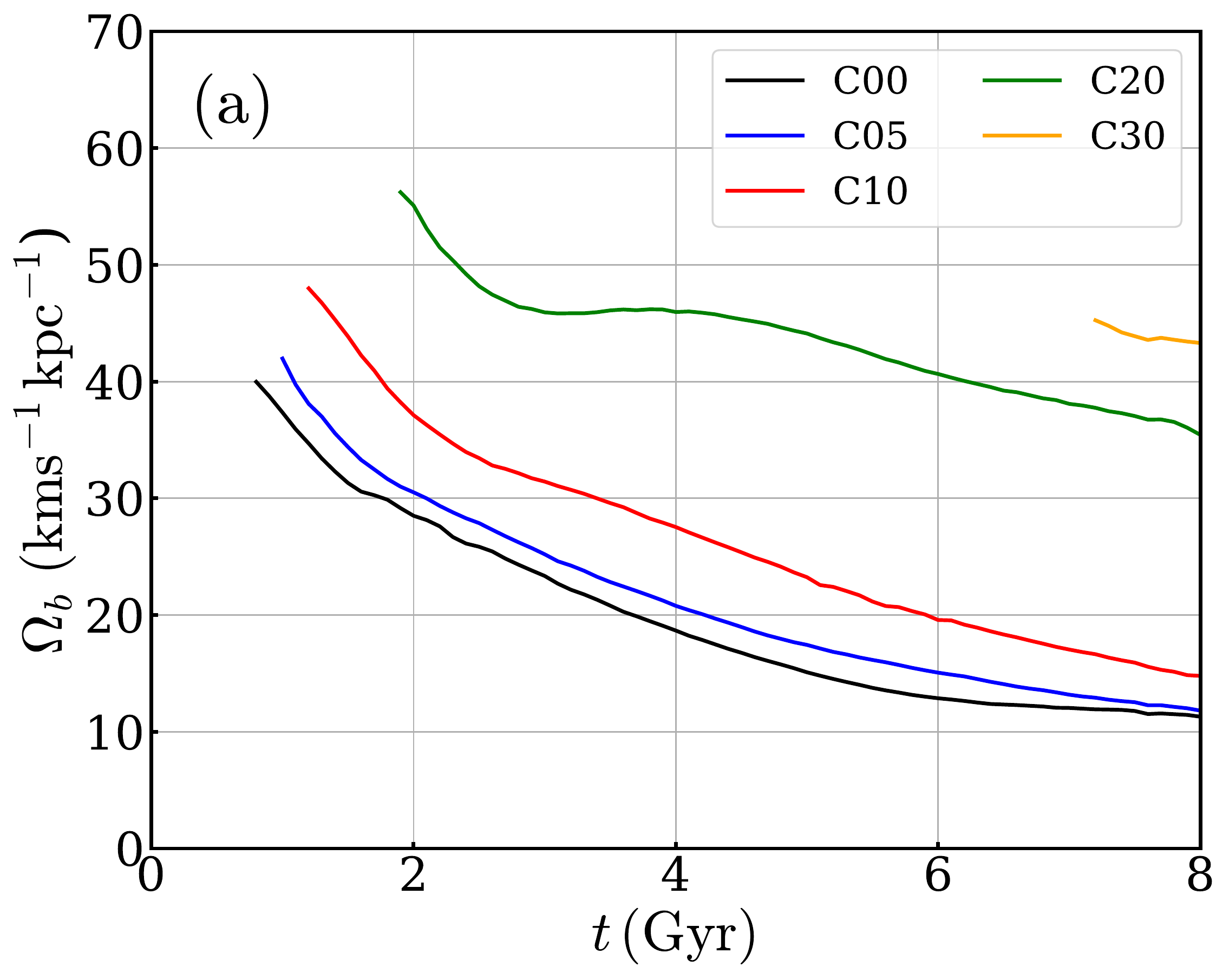}
\plotone{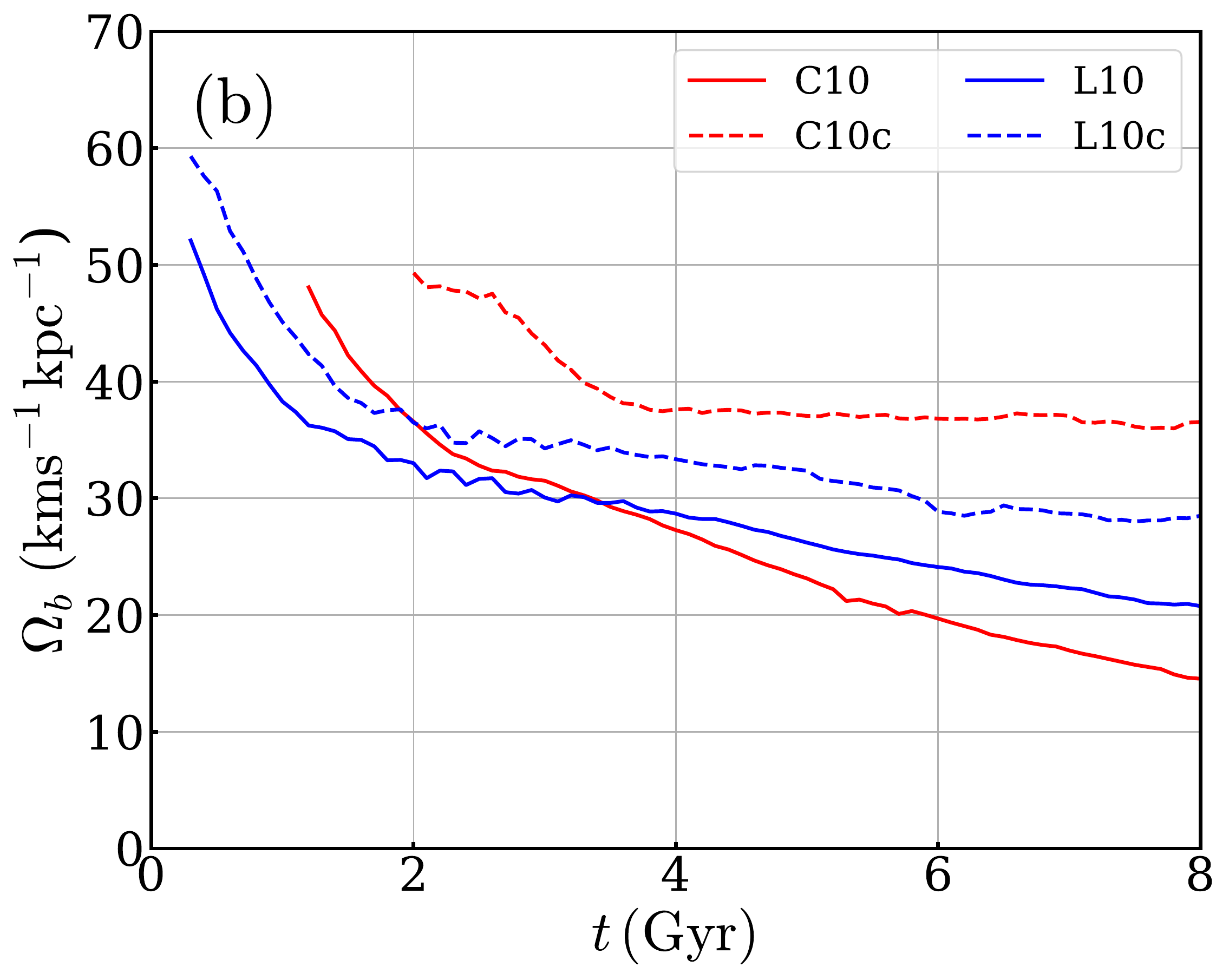}
\caption{Temporal changes of the bar pattern speed $\Omega_b$ for the bar-forming models shown in \autoref{fig:A2_LC}. In all models, $\Omega_b$ decreases over time due to angular momentum transfer from a bar to both halo and bulge.
\label{fig:omegaPL}}
\end{figure}

To calculate the bar pattern speed $\Omega_{b}$, we use two methods: (1) the cross-correlation of the disk surface density in the annular regions with width $\Delta R=0.1\kpc$ at $R=2\kpc$ where most bars attain their maximum strength and (2) the temporal rate of changes in the position angle $\psi$, i.e., $\Omega_b=d\psi/dt|_{R=2\kpc}$. We check that the two methods yield the pattern speeds that agree within $\sim1\%$.
\autoref{fig:omegaPL} plots evolution of the bar pattern speeds for selected models. 
The initial bar pattern speed depends on the bulge mass in such a way that a more massive bulge tends to have larger $\Omega_b$ \citep{kd18,kd19}. In all models, a bar becomes slower over time due to the transfer of angular momentum to both halo and bulge.

\begin{figure}[t] 
\centering
\plotone{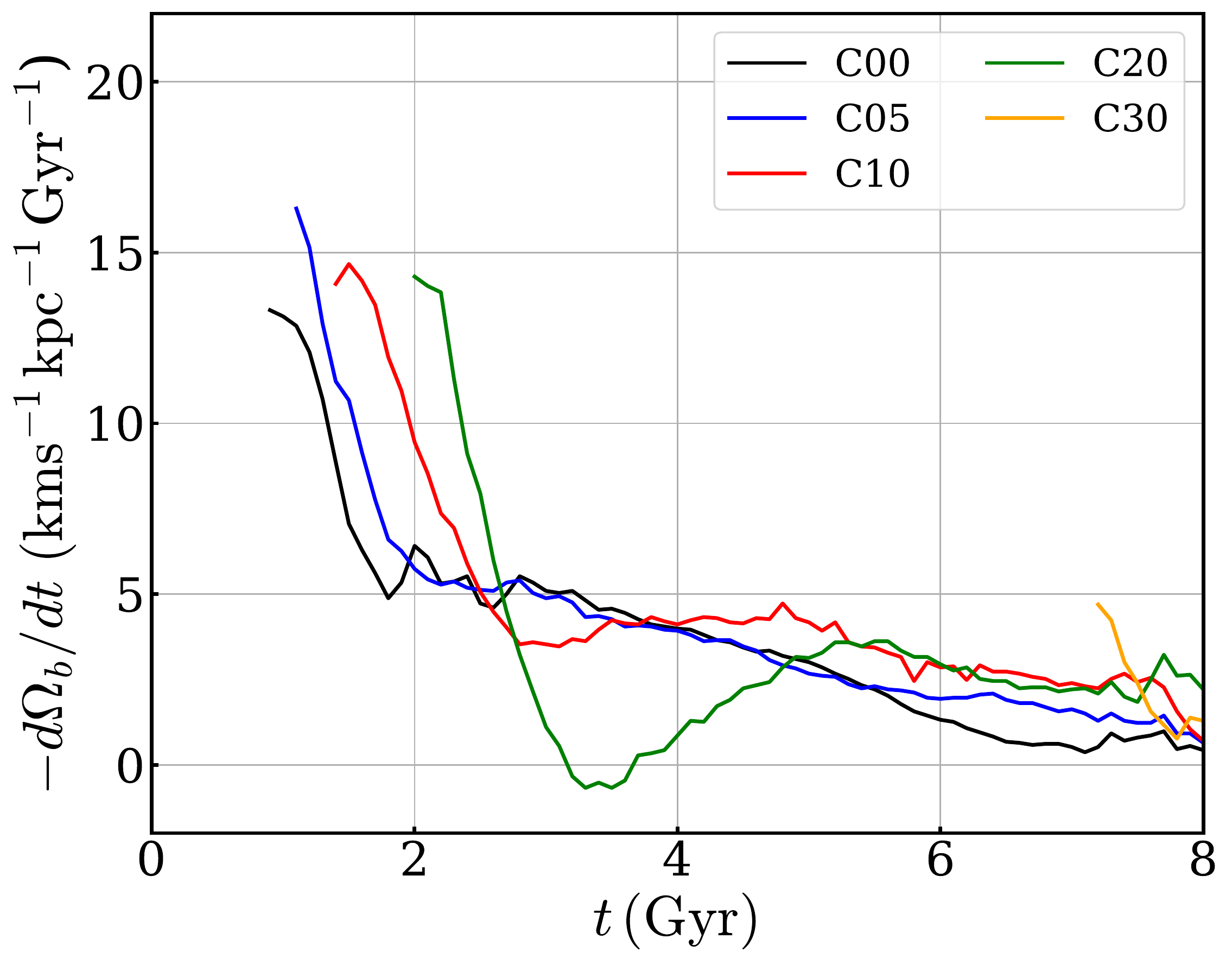}
\caption{Temporal variations of the slow-down rate of the bar pattern speed for the models shown in \autoref{fig:omegaPL}($a$). 
\label{fig:dOdt}}
\end{figure}

\autoref{fig:dOdt} plots the temporal changes in the bar slow-down rate, $-d\Omega_b/dt$, for the models shown in \autoref{fig:omegaPL}($a$), showing that there is no systematic dependence between the bar slow-down rate and the bulge mass. This result is different from \citet{kd19} who found that the rate is higher for a more massive bulge. The discrepancy may be due to the differences in the bulge (and halo) compactness. The models considered by \citet{kd19} have $R_b/R_d \leq 0.18$ with $R_b$ being the half-mass bulge radius, which is more compact than our models in the \texttt{C} series that have $R_b/R_d=(1+\sqrt{2})a_b/R_d=0.32$. Figure 12 of \citet{kd18} shows no systematic trend between the bar slow-down rate and the bulge mass for models with $0.43 < R_b/R_d < 0.47$. This suggests the bulge should be sufficiently compact to control the temporal evolution of the bar pattern speed. In our models with less compact bulge than in \citet{kd19}, angular momentum is predominantly absorbed by the halo (see \autoref{fig:Lz}).

\subsection{Bar Length} \label{subsec:Lbar}

\begin{figure}[t]
\centering
\epsscale{1.0}\plotone{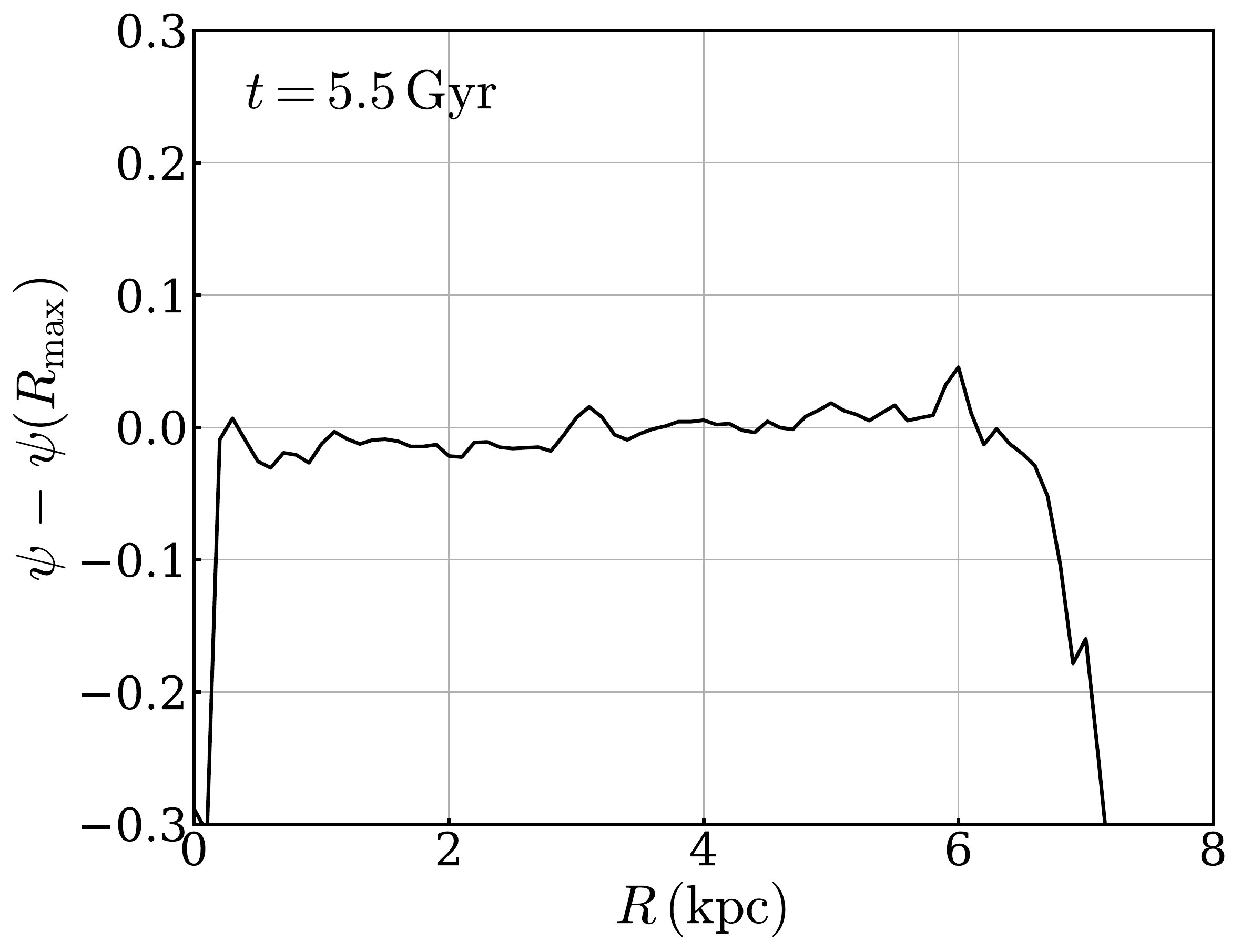}
\caption{Radial distribution of  $\psi-\psi(R_{\rm max})$ of the $m=2$ mode in the disk of model \texttt{C10} at $t= 5.5\Gyr$. Here, $R_\textrm{max}$ denotes the radius where $A_2/A_0$ is maximized. 
The bar has a length $R_b=6.8\kpc$ at this time.
\label{fig:Psi}}
\end{figure}

One can use the position angle $\psi(R)$ defined in \cref{e:posang} to measure the bar length (e.g., \citealt{athmis02,scaath12}).
\autoref{fig:Psi} plots the radial distribution of the position angle of the $m=2$ mode in the disk of model \texttt{C10} at $t= 5.5\Gyr$. Note that $\psi(R)$ that remains more or less constant at small $R$ changes abruptly at $R\gtrsim 6.8\kpc$, indicating that the bar has a semi-major axis $R_b = 6.8 \kpc$ at this time.

\begin{figure}[t] 
\centering
\epsscale{0.95}\plotone{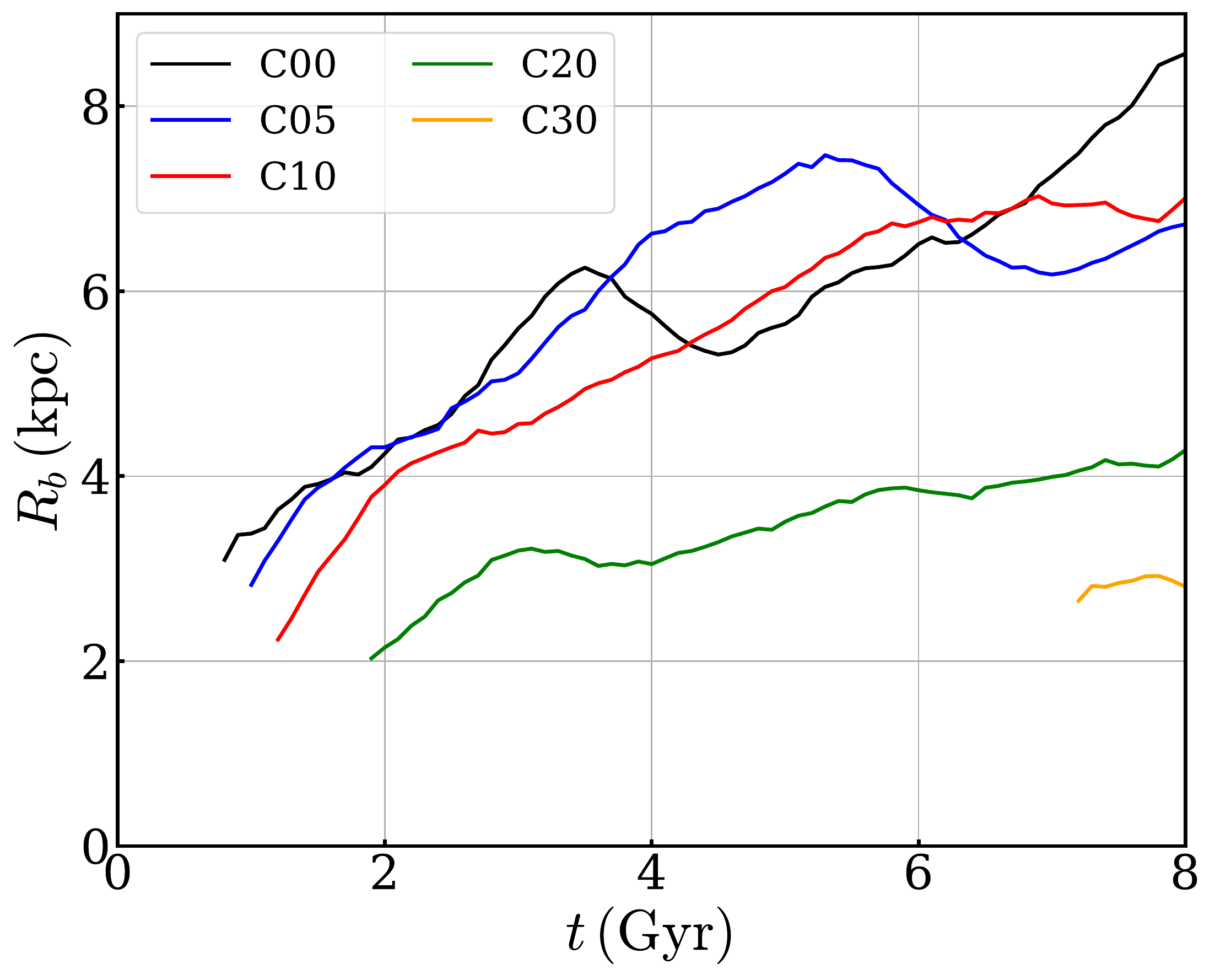}
\epsscale{0.95}\plotone{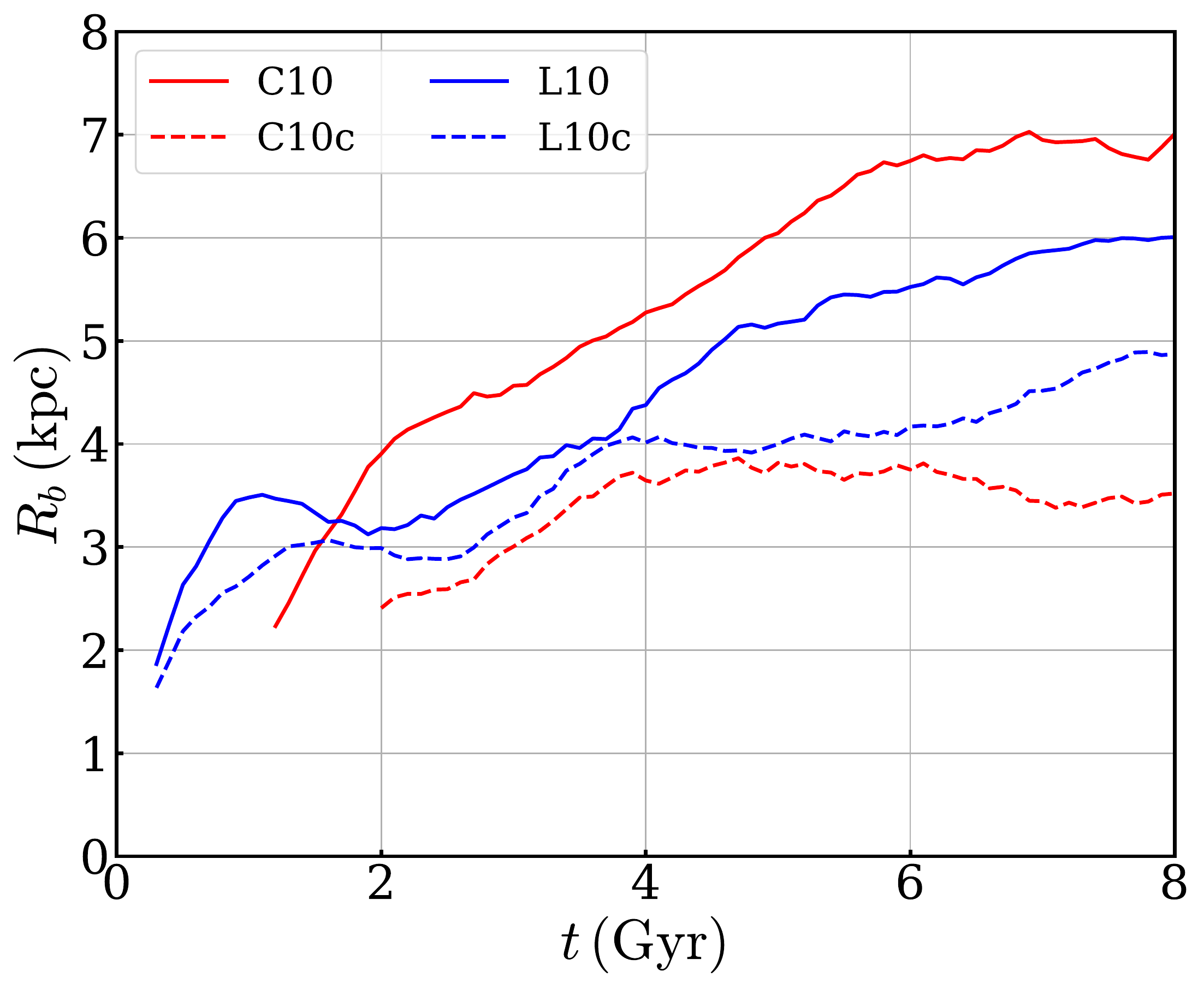}
\caption{Temporal changes of the bar length $R_b$ for the models shown in \autoref{fig:omegaPL}.
\label{fig:Lbar}}
\end{figure}

\autoref{fig:Lbar} plots temporal changes of $R_b$ for the models shown in \autoref{fig:omegaPL}. First of all, bars are longer in models with a less massive and/or less compact bulge since these allow stronger swing amplifications. Overall, the bar length in our models increases with time, expedited by angular momentum exchange with the halo and bulge \citep{ath03}. The increasing rate of the bar length is lower in models with more massive and compact bulge. We note that the decrease of the bar length at $t\sim3.8\Gyr$ in model \texttt{C00}, $t\sim5.3\Gyr$ in model \texttt{C05},  and $t\sim3.3\Gyr$ \texttt{C20} is caused by the interactions with surrounding spiral arms (or an inner ring) which tend to shorten the bar by perturbing particles on outer $x_1$ orbits. In model \texttt{L10}, outer spiral arms are in phase with the bar at $t\sim 1\Gyr$,  making $R_b$ longer than the true bar length temporarily.

\begin{figure}[t] 
\centering
\plotone{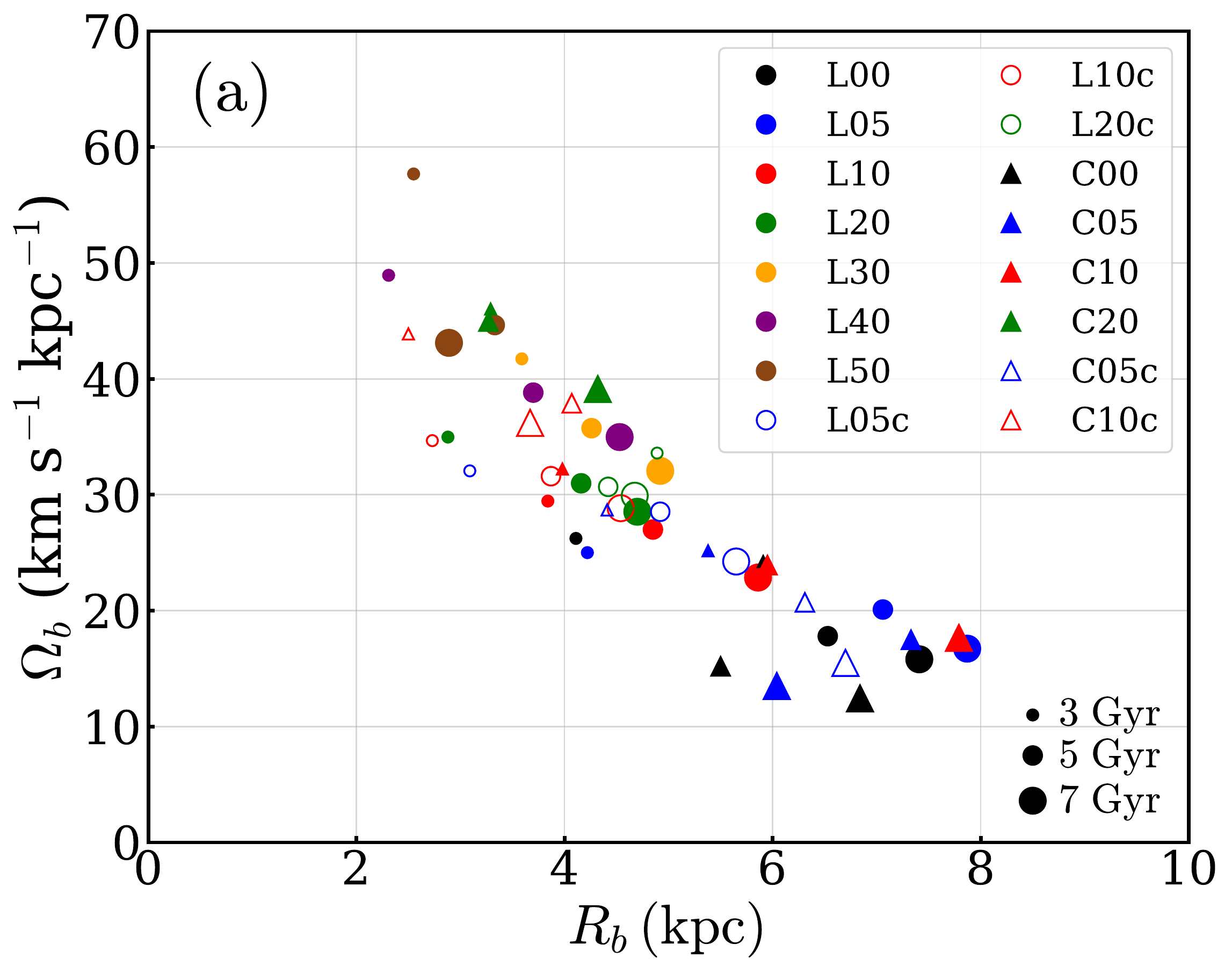}
\plotone{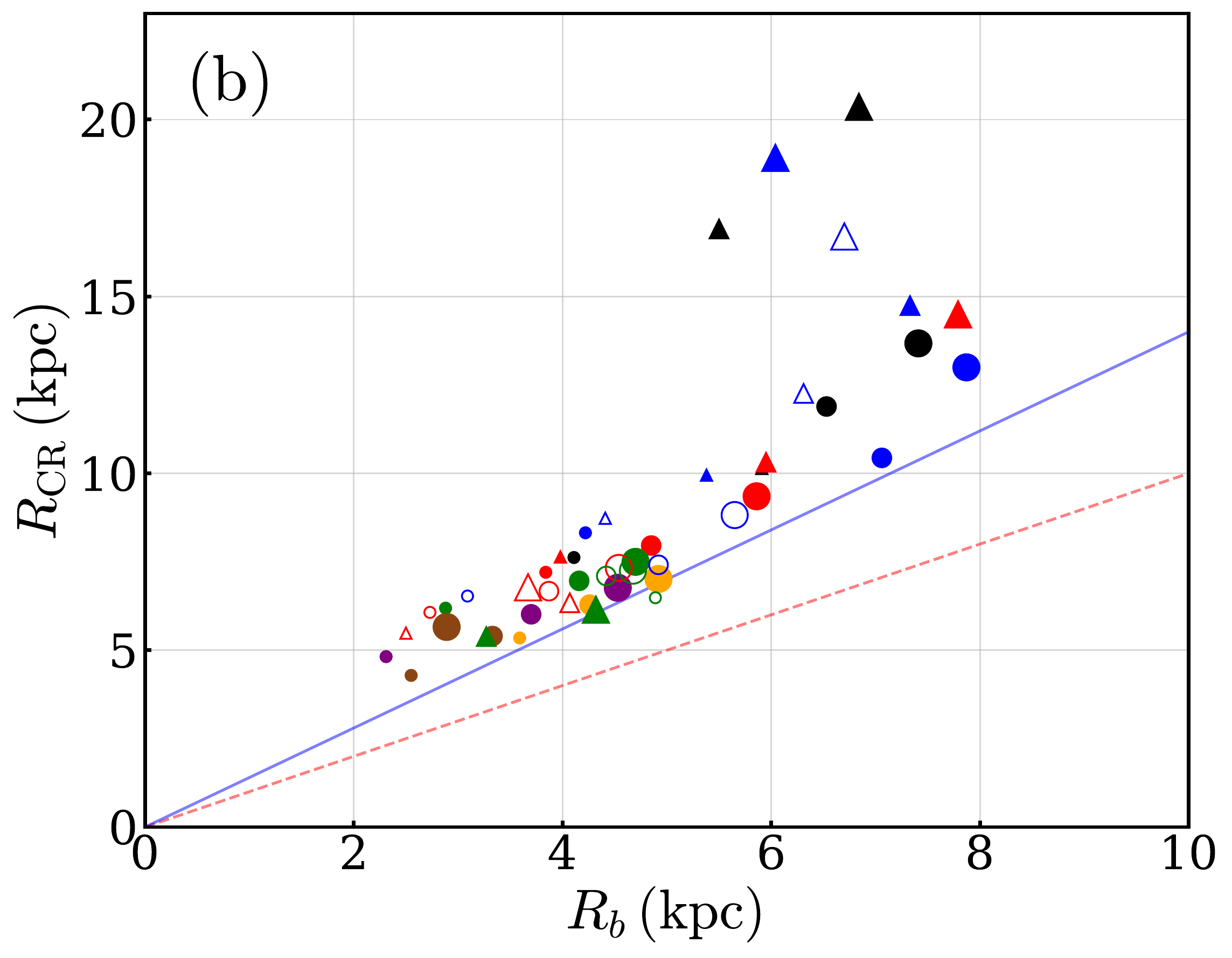}
\caption{
Relationship ($a$) between $\Omega_b$ and $R_b$ 
and ($b$) between $R_\text{CR}$ and $R_b$ for all models that form a bar. The marker sizes correspond to the simulation times. The red dashed and blue solid lines in ($b$) draw $\mathcal{R}\equiv R_\text{CR}/R_{b}=1.0$ and 1.4, respectively, indicating that all bars in our models are slow with $\mathcal{R}>1.4$. 
\label{fig:Rparam}}
\end{figure}

\autoref{fig:Rparam} plots the dependence of the bar pattern speed $\Omega_b$ and the corotation radius $R_\text{CR}$ on the bar length $R_b$ in all models that form a bar, with the symbol size representing the simulation time. In general, longer bars tend to be slower. The ratio $\mathcal{R} = R_{\rm CR}/R_b$ is useful to classify slow or fast bars: bars with $\mathcal{R} > 1.4$ are considered slow, while those with $\mathcal{R} <  1.4$ are termed fast bars.  Models with a massive and compact bulge have larger $\mathcal{R}$ since they have short bars compared to those with a less compact bulge. Note that all bars are slow rotators for almost all time.

\section{Discussion} \label{sec:discussion}

In the preceding section, we have shown that models with a massive and compact bulge and a concentrated halo are less likely to form a bar. 
In this section we compare our numerical results with the previous bar formation conditions mentioned in \autoref{sec:intro}.  We then propose a new two-parameter condition that is consistent with the theory of bar formation. We also use our numerical results to indirectly measure the mass of the classical bulge in the Milky Way.  

\begin{figure}[t] 
\centering
\plotone{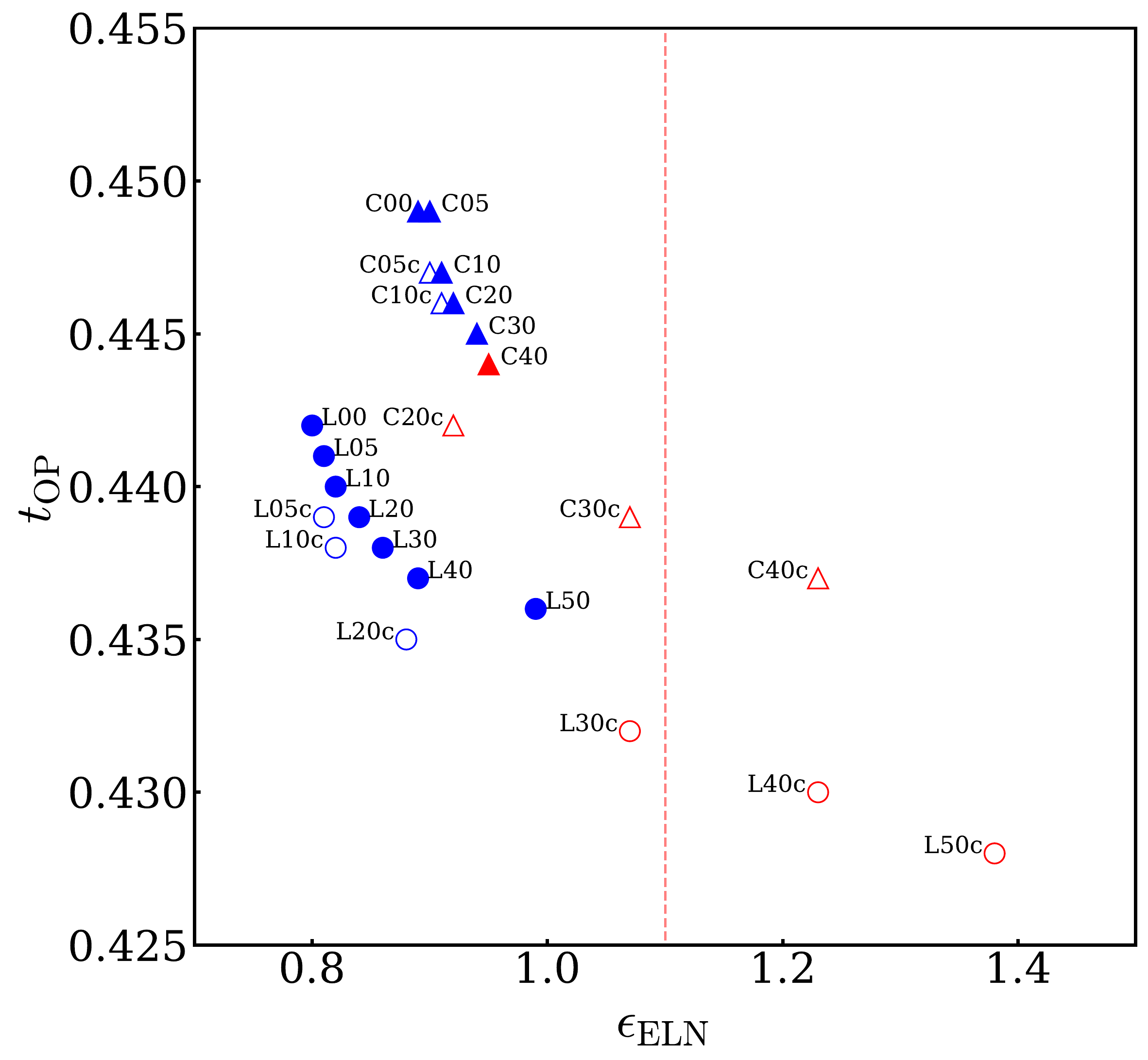}
\caption{Simulation outcomes in the $t_\text{OP}$--$\epsilon_\text{ELN}$ plane. The vertical dashed line marks $\epsilon_\text{ELN}=1.1$ (\autoref{e:eELN}).
Blue symbols denote unstable models for bar formation, while red symbols are for stable models. Circles and triangles are for models in the $\tt L$ and $\tt C$ series, respectively. Open and filled symbols correspond to models with compact and less compact bulges, respectively.
\label{fig:ref1}}
\end{figure}

\begin{figure}[t] 
\centering
\plotone{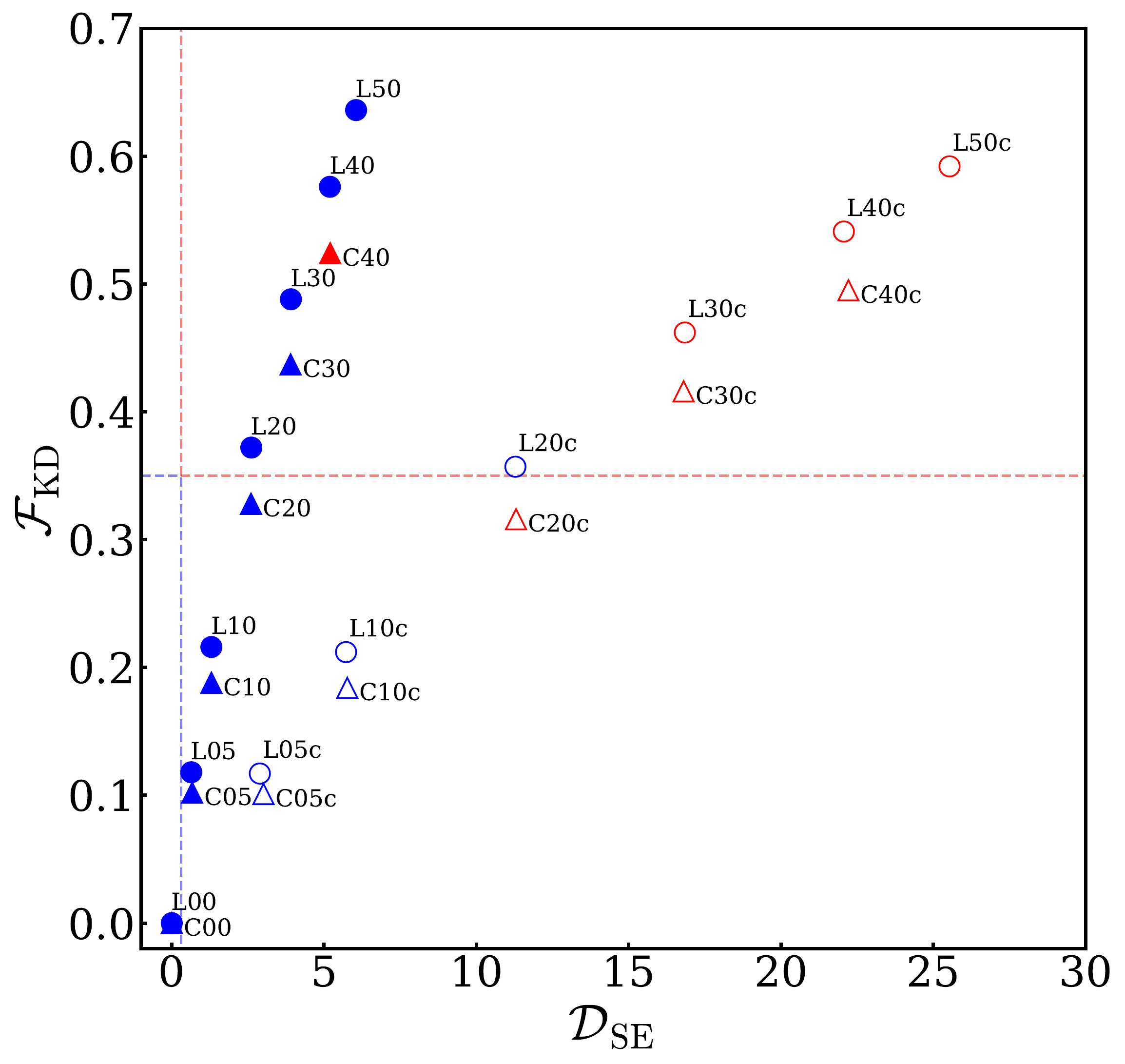}
\caption{Same as \autoref{fig:ref1} but in the
$\mathcal{F}_\text{KD}$--$\mathcal{D}_\text{SE}$ plane. 
The horizontal and vertical dashed lines draw $\mathcal{F}_\text{KD}=0.35$ and $\mathcal{D}_\text{SE}=1/\sqrt{10}$ (see  \cref{e:FKD,e:SE}), respectively. \label{fig:ref2}}
\end{figure}

\subsection{Criteria for Bar Formation} \label{subsec:discuss1}

\autoref{fig:ref1} plots the simulation outcomes in the $t_\text{OP}$--$\epsilon_\text{ELN}$ plane, with the blue and red symbols representing unstable and stable models to bar formation, respectively: the values of $t_\text{OP}$ and $\epsilon_\text{ELN}$ of each model are listed in Columns (7) and (8) of  \autoref{tbl:model}. 
Circles and triangles mark the models in the $\tt L$ and $\tt C$ series, respectively, with the open (filled) symbols corresponding to the compact (less compact) bulges.  While all the models have $t_\text{OP}>0.42$, some of them do not evolve to form a bar, suggesting that $t_\text{OP}$ is not a good indicator of the disk stability against bar formation. This is most likely because \citet{onp73} employed models with a fixed halo, neglecting halo-disk interactions which are crucial for the bar growth. \citet{se18} also noted that the initial value of $t_\text{OP}$ cannot determine whether a bar forms or not.

The abscissa of \autoref{fig:ref1} shows that all the bar-forming models satisfy the ELN criterion (\autoref{e:eELN}). However, some galaxies with a massive bulge under a concentrated halo remain stable even with $\epsilon_\text{ELN}<1.1$. The discrepancy between the ELN criterion and our results is because it, based on 2D thin-disk models with a fixed halo, does not capture the disk-halo interactions (e.g., \citealt{ath08,fujii18}). Analyses of galaxies in recent simulations for cosmological galaxy formation such as EAGLE and IllustrisTNG, etc.\ have also found that $\epsilon_\text{ELN}$ is incomplete to predict whether galaxies formed are barred or not \citep{yu15, algorry17,mario22,izq22}. 

\autoref{fig:ref2} plots our results in the $\mathcal{F}_\text{KD}$--$\mathcal{D}_\text{SE}$ plane, with the blue and and symbols corresponding to the 
unstable and stable models, respectively: the values of $\mathcal{F}_\text{KD}$ and $\mathcal{D}_\text{SE}$ of each model are given in Columns (9) and (10) of \autoref{tbl:model}.
The ordinate of \autoref{fig:ref2} shows that  \cref{e:FKD} is overall consistent with the simulation results for the models in the {\tt C} series, although it fails for the models in the {\tt L} series: some models with a massive bulge under a less concentrate halo form a bar even with $\mathcal{F}_\text{KD}>0.35$. In fact, all the models in \citet{kd18} belong to our {\tt C} series, so that their criterion is unable to predict bar formation in models with less concentrated halos.\footnote{The halos employed in \citet{kd18} have the scale radius of $a_h=17.88\kpc$ for the MA models and $a_h=25.54\kpc$ for the MB models (S.~K.\ Kataria, 2022, private communication), which are smaller than $a_h=30\kpc$ for the models in our {\tt C} series.}

\citet{se18} found that a compact bulge suppresses feedback loops by making the ILR strong. According to \cref{e:SE} for bar formation, all of our models except models \texttt{C00} and \texttt{L00} with no bulge should not form a bar. However, the abscissa of \autoref{fig:ref2} shows that most models with $\mathcal{D}_\text{SE} \lesssim (4$--$10$) are unstable  to bar formation. It is unclear why our results are so different from \cref{e:SE}, but the parts of the reason may be that compared with our models, their halos are small in mass with $M_h\sim 4M_d$ and their disks are thin with $z_d\sim 0.02R_d$.

\begin{figure}[t]
 \centering
 \plotone{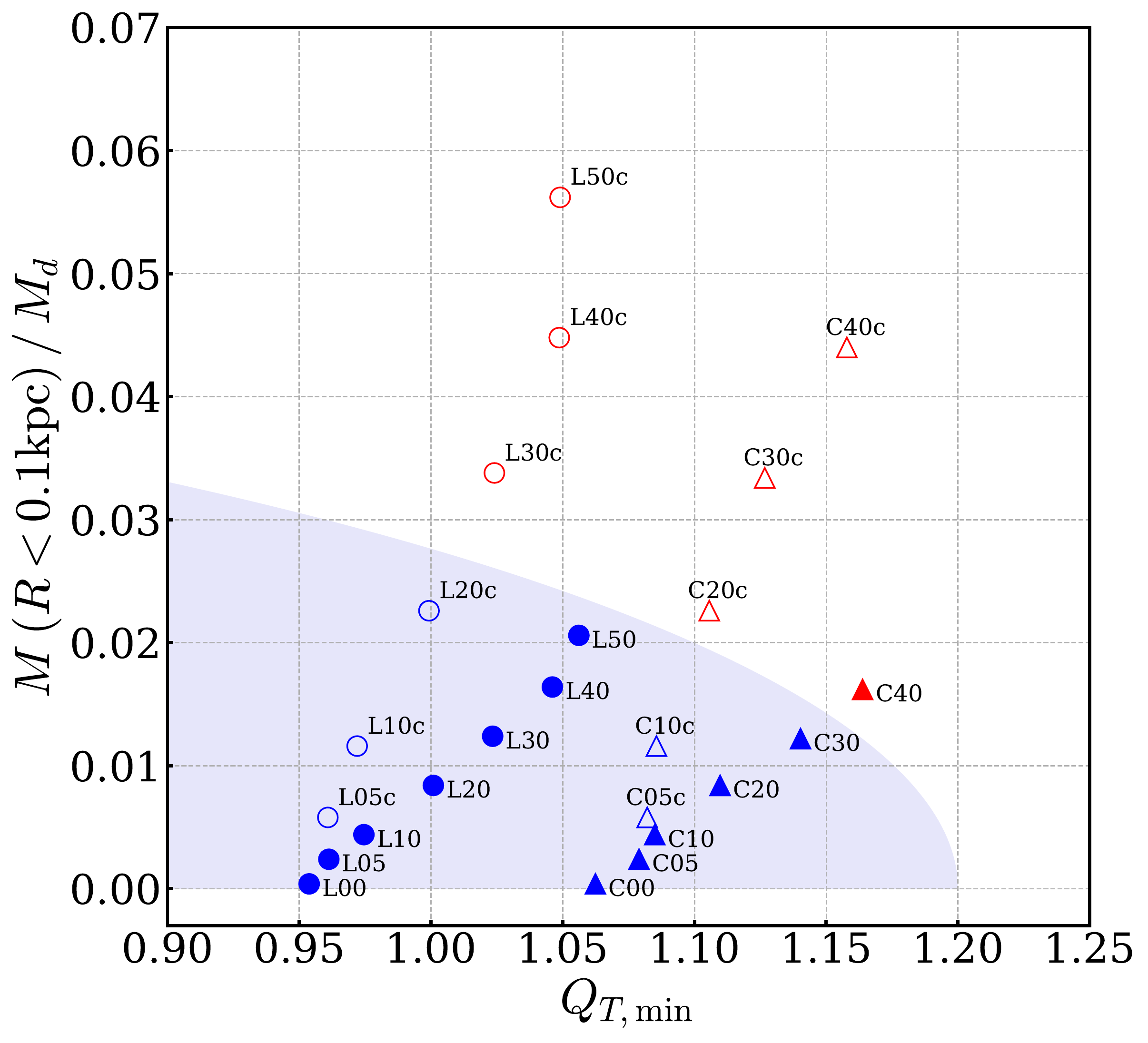}
 \caption{Same as \autoref{fig:ref2} but in the 
$\QTmin$--CMC plane. The shaded regions correspond to \autoref{e:new}, within which all the bar-forming models are located. 
\label{fig:barcri}}
 \end{figure}

As mentioned earlier, bar formation in a disk involves several cycles of swing amplifications and feedback loops. This naturally requires two conditions: (1) the disk should have small $\QTmin$ to be sufficiently susceptible to self-gravitational instability and (2) the ILR should be weak enough for incoming waves pass through the center, which is achieved when the CMC is small. Motivated by these physical considerations, \autoref{fig:barcri} plots the simulation outcomes in the $\QTmin$--CMC plane. Models with more compact bulge and halo have a higher CMC than the less concentrated counterparts with the same $M_b/M_d$. Models with concentrated halos tend to have higher $\QTmin$, although $\QTmin$ is insensitive to the bulge compactness. Note that all the bar-forming models satisfy
\begin{equation}\label{e:new}
    \left(\frac{\QTmin}{1.2}\right)^2 + 
    \left(\frac{\text{CMC}}{0.05}\right)^2 < 1,
\end{equation}
marked as a shade in \autoref{fig:barcri}. In models with $\QTmin$ or CMC larger than \autoref{e:new} implies, swing amplifications with suppressed feedback loops are not strong enough to promote bar formation: these models end up with only weak spiral arms in outer disks (see \autoref{fig:snapall}).

Failure of \cref{e:FKD,e:SE} as a criterion for bar formation is because they account only for a bulge in setting the ILR. However, our results show that not only the bulge mass but also the halo mass in the galaxy center are important in determining the strength of the ILR.

\subsection{Fast or Slow Bars} 

The fact that all bars in our models are slow is consistent with the results of \citet{roshan21} who found that bars formed in cosmological hydrodynamical simulations are preferentially slow, with the mean value of $\mathcal{R}\sim1.9$--$3.0$. However, \citet{cuo20} showed that most observed bars in 77 nearly galaxies are fast, with a mean value of $\mathcal{R}\sim0.92$. What causes the discrepancy in the bar properties between observations and simulations is a challenging question. \citet{frank22} argued that the  discrepancy arises not because the simulated bars are too slow but because they are too short.  

There is a large room for improvement in both simulations and observations for more reliable comparisons. In simulations, our isolated-galaxy models need to be more realistic by including a gaseous component, star formation, halo spin, etc., which may affect the bar pattern speeds significantly. Cosmological simulations still suffer from issues such as insufficient resolution and calibration of feedback from star formation and active galactic nuclei. In observations, the often-used Tremaine-Weinberg method in measuring the bar pattern speeds depends critically on the assumptions that galaxies are in a steady state and that there is a well-defined pattern \citep{tre84}, the validity of which is not always guaranteed. In addition, the bar length depends considerably on the measurement methods such as Fourier analysis, force ratio, ellipse fitting, etc. \citep{Lee22}. Theoretically, it is impossible to have a long-lived, quasi-steady bar with $\mathcal{R}<1$ since the bar-supporting $x_1$ orbits exist only inside the corotation radius (e.g., \citealt{cont80,cont89,bnt08}).

\subsection{Classical Bulge of the Milky Way} \label{subsec:discuss2}

The Milky Way is a barred galaxy dominated by a B/P bulge (e.g., \citealt{dwe95,mar11,nes13}).  
Some early studies reported that the bar in the Milky Way is fast and short, with $50 <\Omega_b <60 \,\rm km\,s^{-1}\,kpc^{-1}$ and $R_b \sim 3 \kpc$ \citep{fux99,deba02,bissan03,frag19,deh00}, while 
recent studies suggested that it is rather slow and long, with $33 <\Omega_b <45\, \rm km\,s^{-1}\,kpc^{-1}$ and $R_b \sim 4.5$--$5 \kpc$ \citep{weg15,sor15,port17,bg16,cla22}. By comparing observed proper motions in the bar and bulge regions with dynamical models, \citet{cla22} most recently reported $\Omega_b = 33.29 \pm 1.81\, \rm km\,s^{-1}\,kpc^{-1}$, placing the corotation resonance at $R_\text{CR} \sim 5$--$7\kpc$. 

Using our numerical results, we attempt to constrain the mass of the classical bulge of the Milky Way. As \cref{fig:omegaPL,fig:Lbar} show, the bar in model \texttt{C10} has $\Omega_b\sim30$--$35\, \rm km\,s^{-1}\,kpc^{-1}$ and $R_b\sim4.5$--$5\kpc$ at $t=2.5$--$3.5\Gyr$, which are well matched to the observed properties of the Milky-Way bar. Model \texttt{C20} produces a bar with $\Omega_b\sim36\, \rm km\,s^{-1}\,kpc^{-1}$ and $R_b\sim4.2\kpc$ at $t=8\Gyr$. The bars in models \texttt{C00} and \texttt{C05} have a length of $R_b\sim 5\kpc$ at $t\sim2.5\Gyr$, but their pattern speeds are smaller than $30\, \rm km\,s^{-1}\,kpc^{-1}$.
These results suggest that the Milky may possess a classical bulge with mass $\sim10$--$20$\% of the disk mass. This is consistent with the claim of \citet{shen10} that the classical bulge of the Milky Way should be less than 25\% of the disk mass to be fitted well with the observed stellar kinematics (see also \citealt{dim15}). If the age of the Milky-Way bar is $\sim3\Gyr$, as proposed by \citet{cole02} based on the ages of infrared carbon stars,  the bar in model {\tt C10} best represents the Milky Way bar. If it is instead $\sim 8\Gyr$ old, as proposed by \citet{bovy19} based on the kinematic analyses of APOGEE and \emph{Gaia} data, it would be better described by the bar in model {\tt C20}.

\section{Conclusions}\label{sec:summary}

We have presented the results of $N$-body simulations to study the effects of spherical components including a classical bulge and a dark halo on the formation and evolution of a bar. For this, we have constructed 3D galaxy models with physical conditions similar to the Milky Way and varied the bulge-to-disk mass ratio as well as the compactness of the halo and bulge components, while fixing the disk and halo masses. Our main conclusions are highlighted below.

\begin{enumerate}
\item \emph{Bar Properties} -- The presence of a massive bulge delays the bar formation.  A bar forms later and weaker in models with a more massive and compact bulge and under a more concentrated halo. Bars are shorter and thus rotate faster in models with more massive and compact bulges. Angular momentum transfer from a bar to both halo and bulge makes the bar slower and longer over time, although most of the angular momentum lost by the bar is absorbed by the halo. All the bars in our models are slow rotators with  $\mathcal{R}=R_\text{CR}/R_b>1.4$.

\item \emph{B/P Bulge and Buckling Instability} -- All the models that form a bar undergo disk thickening and eventually develop a B/P bulge. In all models with a bulge, this proceeds secularly as the bulge tends to suppress the bar formation. However, two models (\texttt{L00} and \texttt{C00}) with no bulge experience buckling instability at $t\sim 5\Gyr$ during which the bar thickens rapidly. The buckling instability occurs when $\sigma_z/\sigma_R$ is kept below $\sim 0.6$ and involves asymmetric density distribution of the disk across the $z=0$ plane. 

\item \emph{Conditions for Bar Formation} -- Our numerical results for bar formation are not well explained by the singe-parameter criteria proposed by the previous studies. We instead find that the bar formation in our galaxy models need to satisfy \cref{e:new}. In models with larger $\QTmin$ or larger CMC, the growth of perturbations via swing amplifications combined with feedback loops is too weak to produce bar-supporting $x_1$ orbits.

\item \emph{Classical Bulge of the Milky Way} -- Among our models, the bar at $t\sim 2.5$--$3.5\Gyr$ in model \texttt{C10} or at $t\sim 8\Gyr$ in model \texttt{C20} is matched well with the observed ranges of the bar length and pattern speed in the Milky Way. This suggests that the Milky Way is most likely to possess a classical bulge with mass  $\sim10$--$20$\% of the disk mass. 

\end{enumerate}

\section*{acknowledgments}
We are grateful to the referee, Dr.\ Sandeep Kumar Katari, for an insightful report. This work was supported by the grants of National Research Foundation of Korea (2020R1A4A2002885 and 2022R1A2C1004810). Computational resources for this project were provided by the Supercomputing Center/Korea Institute of Science and Technology Information with supercomputing resources including technical support (KSC-2022-CRE-0017).

\bibliography{1ms}

\end{document}